\documentclass[reprint,prb,superscriptaddress]{revtex4-1}
\usepackage{amsmath,amssymb,amsfonts,bm}
\usepackage{H1} 

\DeclareMathOperator{\lcm}{lcm}

\newcommand{\zt}{\mathbb{Z}_2}
\newcommand{\ts}{\tilde{\sigma}}
\newcommand{\ta}{\tilde{\tau}}
\newcommand{\on}{\overline{n}}
\begin{document}
\title{Walker-Wang models and axion electrodynamics }

\author{C.W.~von Keyserlingk}
\affiliation{Rudolf Peierls Centre for Theoretical Physics, 1 Keble Road, Oxford,
OX1 3NP, United Kingdom}
\affiliation{Princeton Center for Theoretical Science, Princeton University, Princeton, New Jersey 08544, USA}
\author{F.J.~Burnell}
\affiliation{Department of Physics and Astronomy, University of Minnesota, Minneapolis, MN 55455, USA}
\date{\today}

\begin{abstract}

We connect a family of gauge theories (Maxwell theories with a magnetoelectric coupling $\theta = 2 \pi k, k \in \mathbb{Z}$) to the family of 3D topological lattice models introduced by Walker and Wang.  In particular, we show that the lattice Hamiltonians capture a certain strong-coupling limit of these gauge theories, in which the system enters a gapped (confined) phase.  
We discuss the relationship between the topological order exhibited by certain of these lattice Hamiltonians and the characteristic electromagnetic response of the symmetry-protected bosonic topological insulator.
\end{abstract}


\maketitle

\section{Introduction}
An exciting recent advance in our understanding of phases of matter has been the discovery and classification of interacting symmetry-protected (or SPT) phases of bosons\cite{Chen11}.  Like the more familiar fermionic topological band insulator, these phases are ``trivial" in the bulk (meaning that they have only short-ranged entanglement, and in particular no topological order), but have surfaces that necessarily either (1) break the symmetry; (2) are gapless; or (3) are topologically ordered\cite{Vishwanath13}.  In other words, the surface cannot be trivially gapped   while respecting all the symmetries. 
This occurs  when the surface state realizes symmetry in a way that is ``impossible" (i.e., anomalous) in a purely 2D system.  

An important part of the effort to understand these new phases has been the identification of model Hamiltonians that can realize them.  
Though several approaches are possible\cite{Chen11,SenthilLayers,ChenDecorated}, one approach is to use a family of exactly solvable lattice models proposed by Walker and Wang\cite{Walker12} to obtain symmetry-protected phases.\cite{Fidkowski13c,Fidkowski13b,Chen14}  Each Walker-Wang (or `WW') Hamiltonian is constructed from a 2D anyon model (or more generally a pre-modular category) $\mac{T}$; at the 2D boundary of the 3D system all of the anyons in $\mac{T}$ appear as deconfined excitations, and the surface state is topologically ordered.  Further, many of these models have the property that all excitations are confined in the bulk, which is therefore trivial.\cite{vonkeyserlingk13a}  

This combination of topologically ordered surface and trivial bulk is not, however, sufficient to guarantee that a given Hamiltonian realizes an SPT phase at zero temperature -- the surface states must also realize a symmetry in a way that would not be possible in a purely 2D system.  Though in some cases\cite{Fidkowski13c,Fidkowski13b} time-reversal symmetry is realized by the Walker-Wang surface states in this way, typically they lack a global symmetry that could lead to symmetry protection.

An interesting example of Walker-Wang Hamiltonians that are not SPT's are those constructed from the $\mathcal{T}=\text{U(1)}_q$ anyon models. These describe the $\nu=1/q$ Bosonic or Fermionic Laughlin states depending on whether $q$ is even or odd.  On the surface, the corresponding WW ground states have the topological order of the relevant Laughlin state. In the bulk, the bosonic/fermionic cases have trivial/$\mathbb{Z}_2^f$ topological order respectively.\cite{vonkeyserlingk13a}  (Here $\mathbb{Z}_2^f$ indicates a system with one non-trivial fermionic charge, which will be deconfined in the bulk). Neither of these families of models are SPT: the fermionic models are not trivial in the bulk, and the bosonic models have surface states that can be realized by a purely 2D (bosonic) system.  In light of this it is somewhat surprising that if certain types of perturbations (corresponding to introducing confined anyons into the bulk) are forbidden, even the bosonic systems are separated from the trivial phase by a (first-order) phase transition.\cite{Burnell13}


In the present work, we will focus on better understanding this family of models, by arguing that they realize a particular (confined) limit of ``axion electrodynamics" -- i.e., of a Maxwell theory with bosonic matter sources (of charge $p$) and a topological $\theta$ term:
\be \label{Lax}
\mac{L} = \frac{1}{4 g^2} F_{\mu \nu} F^{\mu \nu} + \frac{\theta }{32 \pi^2} \epsilon_{\mu \nu \rho \lambda} F^{\mu \nu} F^{\rho \lambda}
\ee
Such field theories are of interest in condensed matter, for instance in interacting systems with strong spin-orbit coupling\cite{Maciejko14}.
We will argue that the Walker-Wang models describe a phase of this field theory in the limit $g^2 \rightarrow \infty$ with $\theta = 2 \pi k, k \in \mathbb{Z}$. When $k$ is even and $p=k$, the axion theory bulk is neither SPT nor long-range entangled, but the surface realizes a $\nu=1/k$ bosonic Laughlin topological order just like the U(1)$_k$ Walker-Wang model.  When $k$ is odd and $p=2k$, the axion theory has $\nu=1/k$ surface states and $ \mathbb{Z}_2^f$ bulk topological order just like the U(1)$_k$ WW model.  This bulk topological order reflects the fact that (unlike for fermionic systems) $\theta = 2 \pi$ is {\it not} equivalent to $\theta =0$ in the Lagrangian \eqnref{Lax}.  (Rather, we expect $\theta$ to be periodic modulo $4 \pi$)\cite{Vishwanath13}.

Though previous works have offered conjectures about the appropriate (topological) field theory for the Walker-Wang models\cite{Kapustin13a,Kapustin13b, Walker12,vonkeyserlingk13a}, the present work provides a rigorous correspondence between certain Walker-Wang models and topological field theories, and very explicitly identifies them with a particular region of the phase diagram of the axion electrodynamics described by Eq. (\ref{Lax}).

Perhaps the most interesting outcome of this correspondence 
 is that it allows us to connect the family of Walker-Wang models constructed from fermionic Laughlin states (with $\mathbb{Z}_2^f$ topological order in the bulk) and the bosonic topological insulator\cite{Chen13,Vishwanath13} (or BTI).   The BTI is a symmetry-protected phase of interacting bosons in which the surface cannot be rendered trivial without breaking a global U(1)$\rtimes\mathbb{Z}_2^T$ symmetry.  The connection is most easily understood using the approach of Metlitski et al.\cite{Metlitski13}, who proposed that Eq. (\ref{Lax})  with $\theta = 2 \pi$ describes a version of the SPT which is `weakly gauged' i.e. the U(1) gauge field is weakly fluctuating.  
 (Unlike the situation for fermions, for bosonic systems $\theta$ is defined only modulo $4 \pi$).  
The Walker-Wang model describes a limit in which this weakly gauged (and compact) U(1) symmetry is taken to strong coupling, resulting in a gapped phase with only time-reversal symmetry in the bulk.  
The bulk $\mathbb{Z}_2^f$ topological order of the Walker-Wang model is a direct consequence of the fact\cite{Metlitski13} that in the bosonic topological insulator the magnetic monopole (which one expects to proliferate in the strong-coupling limit, leading to confinement) is a fermion.\footnote{This   $\mathbb{Z}_2^f$ topological order at strong coupling has also been pointed out by Ref. \onlinecite{MaxUnpublished1,MaxUnpublished2}.}

This connection between lattice models and the confining phases of the field theory (\ref{Lax}) also gives us insight into the nature of the models constructed from bosonic Laughlin states, which we identify with confined phases at $\theta = 4 \pi n, n \in \mathbb{Z}$.  
In particular, this identification makes clear the sense in which these models are trivial: the bulk phase of matter that they realize is, up to a re-labeling of its excitations, just the ordinary confined phase of Maxwell theory, which is a prototypically ``trivial" phase.  The field theory also suggests a path through the phase diagram that could comprise an adiabatic deformation from the Walker-Wang Hamiltonian to the trivial phase.

The remainder of this work is set out as follows. We begin in \secref{sec:confinedBTI} by discussing qualitatively the gapped phases of field theories of the form (\ref{Lax}). We then identify a limit in which these can be described by a purely topological `$bF+bb$' action; our arguments 
here do not address certain technical challenges with identifying this topological limit of gauge theory for $\theta \neq 0$ on the lattice, and
are partly heuristic. 
 In \secref{WWSec} we also briefly introduce the corresponding Walker-Wang models.  In \secref{sec:DiscretebF+bb} we discuss a lattice version of the $bF+bb$ theory, and show rigorously that its partition function describes a path integral for the Walker-Wang Hamiltonian (in discretized time). In \secref{FTCSec} we focus on the specific example of the bosonic topological insulator and the corresponding Walker-Wang model, discussing common features between the ground states of the lattice model and the non-linear sigma model that describes the BTI.\cite{Xu13}  In \secref{RandomSec} we discuss the boundaries of our confined phases, and the fate of time reversal symmetry in the lattice models. 
 We conclude in \secref{conclusion} with some comments about the generalizability of our results, as well as their implications for the phase diagram of the Walker-Wang lattice models.  In Appendix \ref{SVApp}, we discuss the connection between our lattice models and those proposed for the BTI by Senthil and Vishwanath\cite{Vishwanath13}.

\section{ Gapped phases of the gauged bosonic topological insulator and its cousins}
\label{sec:confinedBTI}

In this section, we will discuss the confined phases of U(1) gauge theory with a topological $\theta$ term $\theta = 2 \pi k$, and introduce the limit in which they are well-described by Abelian Walker-Wang Hamiltonians. (This limit has also recently been considered by Ref.~\onlinecite{Gukov13} in the high energy literature, and Refs.~\onlinecite{MaxUnpublished1,MaxUnpublished2,Ye14, Motrunich14a,Motrunich14b} consider closely related models). For $k$ even, these phases are completely  confining in the bulk; for $k$ odd, they will all have a single species of deconfined emergent fermion, and have $\mathbb{Z}_2$ topological order.  

The starting point for our analysis is compact U(1) gauge theory coupled to charged bosonic matter, in the presence of an axion term \cite{Cardy82a,Cardy82b}; the phase structure in the absence of a theta term (i.e., $\theta=0$) has been extensively studied (see, among others, Refs. \onlinecite{Banks77,StonePRL78,UkawaGuthPRD21}).  Later in \secref{ss:bFbbtoWW:Fieldtheory} we will discuss a related theory on the lattice. In this section, however, we argue qualitatively using the continuum action

\be
\label{eq:effaction}
S =  \int d^4 x \left(  \frac{1}{4g^2} \Gamma ^2_{\mu\nu}  - i  \frac{\theta}{8 \pi^2 } \Gamma \wedge \Gamma +i  p A_\mu n^\mu \right) +  \mac{L}_m\punc{,}
\ee
 where $\Gamma_{\mu \nu} = F_{\mu \nu}+2\pi s_{\mu \nu}$, and $F_{\mu \nu} = \partial_\mu A_\nu - \partial_\nu A_\mu$ is the non-singular part of the field strength, while $s_{\mu \nu}$ accounts for the $2\pi$ flux Dirac strings of monopoles that may also be present since our gauge field is compact.  We have also defined $\Gamma \wedge \Gamma = \frac{1}{4} \epsilon^{\mu \nu \rho \lambda}  \Gamma_{\mu \nu} \Gamma_{\rho \lambda}$.  We use a Euclidean space-time metric, such that $Z = e^{ - S}$; this is responsible for the factors of $i$ in the second and third terms.  $n^\mu$ is the current arising from our bosonic charges, while $\mac{L}_m$ is an action for the matter fields which we take to be
\be\label{eq:Lm}
\mac{L}_m=K_H n^2 \punc{.}
\ee  
This imparts particle world-lines with an energy per unit length of $K_H$, although for the majority of our analysis we work in the limit where $K_H=0$.

 Since the gauge theory is compact, in addition to currents of bosons carrying integral charges, there are currents of monopoles of flux $2\pi \mathbb{Z}$. Throughout this work we will denote a bound state of a charge $n$ boson, and flux $2\pi m$ monopole by $(n,m)$. What difference does the theta term make? When $\theta\neq0$, a magnetic monopole $(0,m)$ couples to the gauge field as though it has electric charge $q_I=\frac{\theta}{2\pi} m$; we say the monopole has an \emph{induced} electric charge $q_I$. More generally, the object $(n,m)$ couples to the gauge theory as though it has total electric charge $q_E = n + \frac{\theta}{2\pi} m$. Correspondingly, in the Hamiltonian formulation of the theory, the theta term modifies the Gauss constraint to  $\partial_i \hat{E}_i = \rho_{b} + \frac{\theta}{2\pi}  \rho_{m}$, so that it depends on the monopole number density $\rho_{m}$ in addition to the boson number density $\rho_{b}$. Here we define the electric flux as
\be
 \hat{E}_i \equiv  \hat{\mathcal{E}}_i - \frac{\theta}{4\pi^2} \hat{B}_i\punc{,} \label{eq:Eflux}
\ee
where $\hat{B}_i \equiv \epsilon_{i j k} \hat{F}_{jk}/2$, and  $-\hat{\mathcal{E}}_{i}\equiv -i \frac{\delta }{\delta \hat{A}_{i}}$ is the momentum canonically conjugate to $\hat{A}_{i}$. In the majority of this chapter, we will assume $ k= \theta/2\pi$ is integer valued. Let us compare the theories at $\theta=2\pi k$ and $\theta=0$. The allowed particles in both theories are $(n,m)$ where $n,m\in \mathbb{Z}$. One naively expects the physics at $\theta=2\pi k$ and $\theta=0$ to be the same: for each particle $(n,m)$ at $\theta=2\pi k$ there is precisely one particle at $\theta=0$, namely $(n+ km,n)$, with the same electric charge and magnetic flux. Thus, one expects the physics at $\theta=2\pi k, 0$ to be the same under the one-to-one relabeling $(n,m) \rightarrow (n+ km,n)$ \cite{Callan76,Jackiw76,Qi08}. 

 However Ref.~\onlinecite{Metlitski13} points out that the physics at $\theta=2\pi k$ and $\theta=0$ is not quite the same if $k$ is odd. For example, $(-1,1)$ and $(0,1)$ have the same electric and magnetic charges at $\theta=2\pi$ and $\theta=0$ respectively -- they are both electrically neutral and carry unit monopole charge at their respective values of $\theta$. However, the bound state $(-1,1)$ is a fermion \cite{Goldhaber76}, while $(0,1)$ is a boson. Therefore the $\theta=2\pi, \theta=0$ theories are physically distinguishable; in the former $(-1,1)$ is electrically neutral and fermionic, while in the latter $(0,1)$ is electrically neutral but bosonic (see \tabref{NotationTable}). This fact was used ingeniously\cite{Metlitski13} to argue that \eqnref{eq:effaction} with $\theta=2\pi$ and small $g^2$ (weak coupling) describes the electromagnetic response of a bosonic topological insulator.

\begin{table}[h]
\begin{tabular}{|c|c|c|c|}
\hline
 name & label & $q_E$ & $q_M$ \\
\hline
 fundamental monopole &$(0,1)$& $\frac{\theta}{2 \pi} $& 1 \\
neutral monopole & $(-\frac{\theta}{2 \pi},1)$ & $0$ & 1 \\
 fundamental charge & $(1,0)$ & 1 & 0 \\
\hline
\end{tabular}
\caption{ \label{NotationTable} Our notation for the point particles in the model \eqnref{eq:effaction}. The first column shows the label that we will use in the text, which denotes the number of bosonic charges and fundamental monopoles in a given object.  This quantity is independent of $\theta$.  The second and third columns show the electric and magnetic charges of each object (the former being $\theta$-dependent).}  
\end{table}

	\subsection{Gapped phases and condensation}\label{ssec:gappedphases}
The question of which, if any, of these particles will condense for a given $(g, \theta, \mac{L}_M)$ was addressed by Cardy and Rabinovici\cite{Cardy82a}; we will briefly review their arguments here.  The interactions  between these point particles can be understood by integrating out $A_\mu$ in \eqnref{eq:effaction},\cite{Cardy82a} to obtain:
  \ba \label{SChargeMon}
S&=& \int d^4 r d^4 r' \left(   \frac{2 \pi^2}{g^2} m_\mu(r) G(r-r') m_\mu (r') \right. \n
&&\left.  \frac{1}{2} p^2 g^2 j_\mu (r) G(r-r') j_\mu(r') \right. \n
&&\left. +2 \pi i p \partial^\nu s_{\mu \nu} (r) n^\mu(r')  G(r-r') \right )  +  \mac{L}_m
\ea
where we have defined the total charge current
\be
j_\mu(r) = n_\mu(r) + \frac{ \theta}{2 \pi p } m_\mu(r)
\ee
and $g_{\mu \nu} G(r-r')$ is the photon propagator in the Feynman gauge.  The first two terms in Eq. (\ref{SChargeMon}) represent the Coulomb repulsion between magnetically and electrically charged objects, respectively.  The total electric current $j^\mu$ contains both the original charge current $n^\mu$ and a term $ \frac{ \theta}{2 \pi p } m^\mu$ proportional to the monopole current, indicating that the monopole now carries an electric charge of $\theta/(2 \pi)$.  The third term in \eqnref{SChargeMon} is a statistical Berry phase interaction between the matter and monopole currents, and which is independent of $\theta$.

 If current loops are relatively dilute, the dominant contribution of the Coulomb force will be to their self-energy, which has a fixed value per unit length of the current loop.   If we fix the total length of a current loop passing through a particular edge, however, there are many different possible loop configurations that enter the partition sum (which we will evaluate in imaginary time).
 Thus a current loop  also effectively has a (fixed) entropy per unit length.  As described in Ref. \onlinecite{Cardy82a}, the competition between self-energy and entropy  leads to an approximate criterion for condensation (for $\mac{L}_m =0$): 
\be \label{CondCond}
\left(\frac{m_\mu^2}{\gamma} + ( n_\mu + \frac{k}{p} m_\mu)^2 \gamma \right ) p < C
\ee
where $\gamma = p g^2/ (2 \pi)$, $k = \frac{\theta}{2 \pi}$,  and $C$ describes the entropy per unit length of the current loop.  

There are four basic things to notice about the phase diagram predicted by Eq. (\ref{CondCond}).  First, at large $g^2$ (i.e. large $\gamma$), the self-energy of any electrically charged object is extremely high, and only neutral objects for which
\be
p n = - k m
\ee
can condense.   For $\theta = 2 \pi k$, $p=k$, the fundamental object satisfying this criterion is the neutral monopole $(-k, 1)$-- a bound state of a charge $-k$ boson and a single monopole with its induced charge of $k$.  (Note that in our notation $(1,0)$ is an object of charge $1$, not an object of charge $p$.)  Because these electrically neutral objects are composites of both fundamental charges and fundamental monopoles, we follow Ref. \onlinecite{Cardy82a} and refer to the gapped phases in which they have condensed as {\it obliquely confined} phases.  

Second,  this neutral monopole exists only if our theory contains {\it dynamical} matter fields of charge $k$.  In the coming sections we will consider the effect of adding non-fluctuating matter fields (of charge $q \in \mathbb{Z}, q< p$) as ``test charges" in our model.  However, in Eq. (\ref{eq:effaction}) we have explicitly included the factor of $p$ because the matter current $n^\mu$ must be summed over in the partition function (i.e., must be able to fluctuate) in order for the neutral monopole to condense.  

Third, if $k$ is even, the neutral monopole is a boson, and can condense; this leads to a phase in which there are no deconfined point particles in the bulk. In this case, we will find it convenient to choose $p=k$, such that only the matter fields that are required to enter the gapped phase are dynamical.  If $k$ is odd, the neutral monopole is a fermion, and must pair in order to condense.  In this case the large $g^{2}$ phase has $\mathbb{Z}_2^f$ topological order, much like a superconductor\cite{Hansson04}.  Here we will find it convenient to choose $p = 2 k$ (again, taking the minimum set of dynamical matter fields required to enter the gapped phase). 

Finally, we can alter the criterion (\ref{CondCond}) somewhat by adding a mass term (contained in $\mac{L}_m$) for our bosons.  Eq.  (\ref{CondCond}) contains only the Coulombic self-energy; a mass term will add to this an energy per unit length for any current loops of $n^\mu$.  Hence by making this mass sufficiently large, we can prevent any particle from condensing at large $g$.

\begin{figure}[ht]
 \begin{center}   
     \includegraphics[width=2.75in]{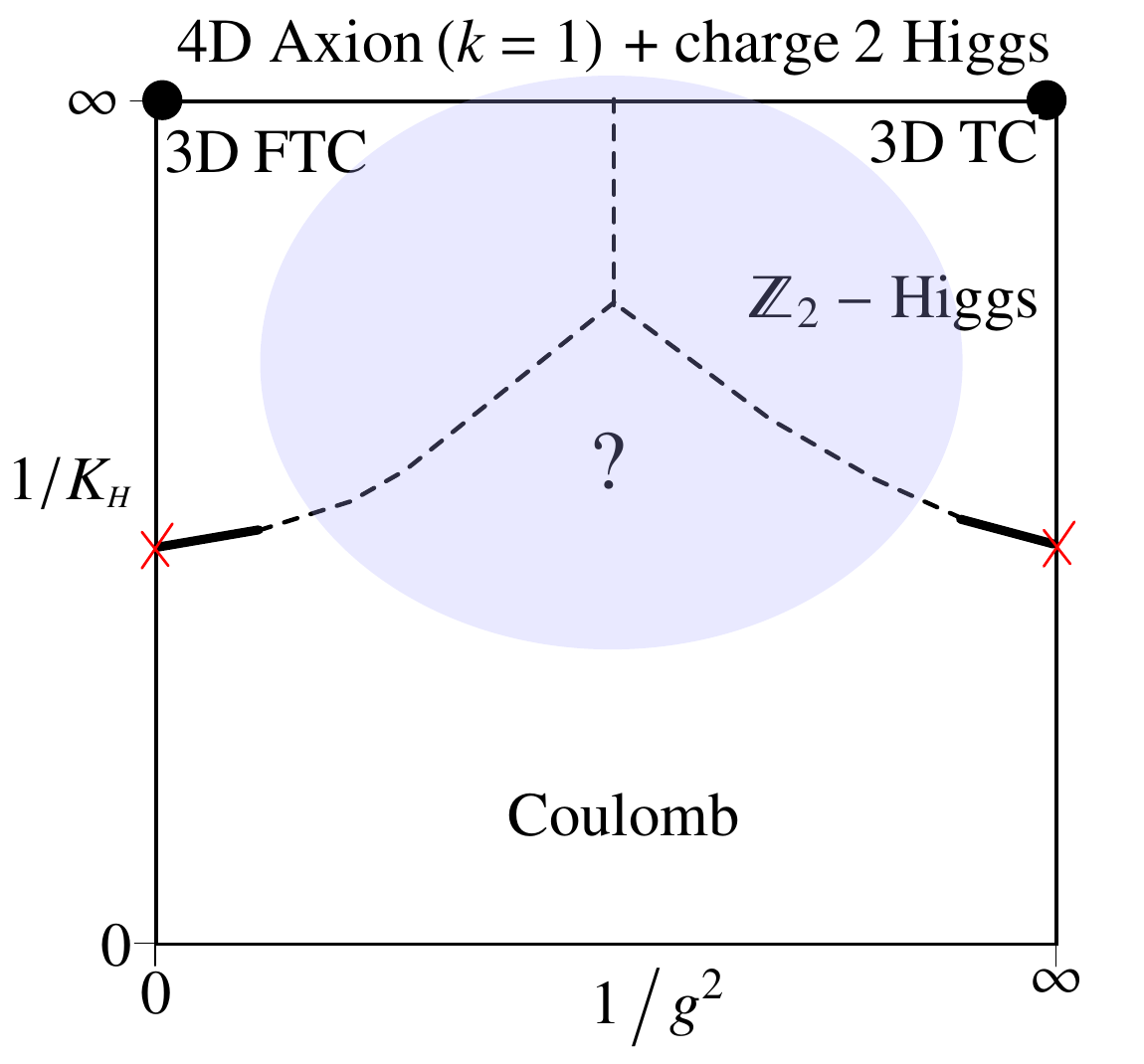}
\caption{Heuristic phase diagram for the action in \eqnref{eq:effaction} for $\theta = 2 \pi, p=2$.  Here $K_H$ is defined in \eqnref{eq:Lm}, and is the energy cost per unit length of the charge $2$ bosonic world-lines.} \label{PhaseFig1}
 \end{center}
 
 \end{figure}
 
It is instructive to consider some examples of the resulting phase diagram.  For $k=1, p=2$ (i.e. $\theta = 2 \pi$, with a dynamical matter field of charge $2$) , the (approximate) phase diagram shown in Fig. \ref{PhaseFig1}.  
The obliquely confined phase at large $g^2$ and the Higgs phase at small $g^2$ are both topologically ordered; however, they have different topological order and are therefore distinct.  We will describe the reasons for this in more detail in Sect. \ref{BTI:gapped}.  Essentially, however, this is because the Higgs phase admits a deconfined charge 1 bosonic excitation, while in the obliquely confined phase we will find a deconfined {\it fermion}.  It is well known that the $p=2$ Higgs phase is described by the 3D Toric code (or $\mathbb{Z}_2$ gauge theory with bosonic charges\cite{Fradkin79}); we will show in the coming sections that the obliquely confined phase at $\theta = 2 \pi$ is described by $\mathbb{Z}_2$ gauge theory with fermionic charges-- i.e., by a fermionic variant of the 3D Toric code.    
 
\begin{figure}[ht]
 \begin{center}
     \includegraphics[width=2.75in]{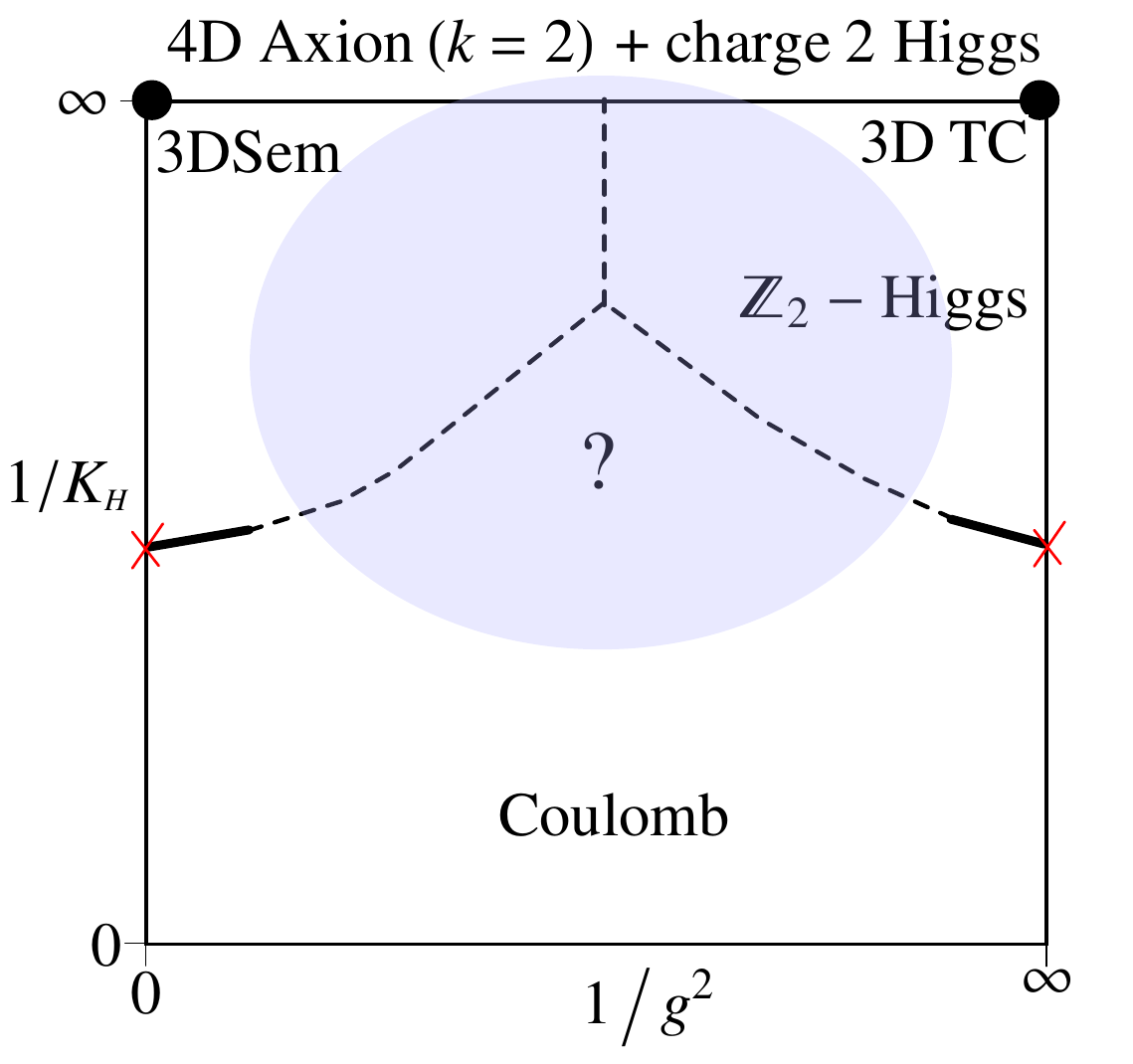}
 
\caption{Heuristic phase diagram for the action in \eqnref{eq:effaction} for $\theta = 4 \pi, p=2$.  Again, $K_H$ is defined in \eqnref{eq:Lm}.}    \label{PhaseFig2}
 \end{center}
 \end{figure}

 For $\theta =  4 \pi $, $p=2$, an approximate phase diagram is shown in Fig. \ref{PhaseFig2}.   The Higgs phase here is identical to that for $\theta = 2 \pi, p=2$: the dynamical charge 2 leads to a gapped phase with the $\mathbb{Z}_2$ topological order of the 3D (bosonic) Toric code.  For $\theta = 4 \pi$ the obliquely confined phase, however, is akin to the usual confined phase at $\theta =0$, with no topological order or deconfined excitations.  
 
 For $k>2$ the Higgs phases are more complicated, with $\mathbb{Z}_k$ ($\mathbb{Z}_{2k}$) topological order for $k$ even (odd).  However, the obliquely confined phases follow the same pattern: if $k$ is odd, the weakly coupled system is a bosonic topological insulator and the obliquely confined phase has $\mathbb{Z}_2^f$ topological order; if $k$ is even the weakly coupled system is a trivial insulator and the obliquely confined phase has no topological order.  


\subsection{The effect of dynamical matter}

 In the remainder of this section, we will discuss the behavior of these various obliquely confined phases in more detail.  
 We argue that with the choice $p=k$ for $k$ even, $p=2k$ for $k$ odd, the spectrum of both confined and deconfined excitations in these models is described by a family of lattice models introduced by Walker and Wang\cite{Walker12}. In \secref{sec:DiscretebF+bb} we will derive a Walker-Wang Hamiltonian starting from a lattice gauge theory related to \eqnref{eq:effaction}, thus making a more concrete connection between Walker-Wang models and the gapped phases of axion gauge theories. 
 
 In this discussion, our choice of $p$ (i.e., the charge of the fundamental dynamical matter field) will play a crucial role.  This is because in a confined phase, 
the energy cost of separating two charges ($\pm q)$ is linear in their separation.  
In the presence of dynamical matter, at large separations this confining interaction will be screened by a second particle-antiparticle pair appearing from the vacuum.  (In other words, in the absence of dynamical matter, the Wilson loop obeys an area law; in the presence of dynamical matter it  obeys a perimeter law\cite{Fradkin79}).   A dynamical matter field of charge $1$ will therefore completely screen all matter fields at  long length scales, effectively eliminating them from the long-wavelength theory-- as is familiar from QCD, where at  low energies there are no bare quarks, only mesons and baryons.  A dynamical matter field of charge $p$, on the other hand, can only screen charge in multiples of $p$.  This leaves $p-1$ possible (non-dynamical) ``test charges" in the theory which cannot be screened and appear as confined excitations in the long-wavelength spectrum.  Closed loops of electric flux can also be screened in multiples of $p$, such that electric flux becomes a $\mathbb{Z}_p$ valued quantity at long distances.  

Of course, it is always possible (and indeed, natural) to introduce dynamical matter fields of charge 1, in which case the confined excitations that we will identify here would disappear from the theory completely.  Here we find that if we do not do this, but instead keep only the minimal set of dynamical fields required to enter the gapped phase, we will recover the confined spectrum of the Walker-Wang model, which can be viewed as describing this particular limit.  We emphasize, however, that the set of confined excitations obtained in this way is {\it not} in any way a universal characteristic of the phase, and that our choice of $p$ is made purely to allow us to make a connection to the Walker-Wang lattice models.  
  

\subsection{Physics in the obliquely confined phase  for $k$  even}\label{BTI:gapped}


We begin by studying the nature of the ``trivial" obliquely confined phase for $k$ even, with $p=k$.  As discussed above (see Table \ref{NotationTable}), we will follow the notation of Cardy and Rabinovici\cite{Cardy82a}, rather than that of Metlitski, Kane, and Fisher\cite{Metlitski13}, and label particles by ($n_0, m_0$) (corresponding to their electric charge due only to the bosonic sources, and their monopole charge).  In this notation, for $\theta = 2 \pi k$ the (electrically neutral) magnetic monopole is $(-k, 1)$.  

If $k$ is an even integer, the neutral monopole $(-k,1)$ is a boson, and the bulk is essentially the usual confined phase of a compact U(1) gauge theory.   Exactly as in the confined phase of compact QED, all excitations with electric charge are confined\cite{Polyakov75,Cardy82a}.  This includes the $(0,1)$ monopole, which has an electric charge of $k$, as well as all charged excitations $(n,0)$.  Screening by the charge $k$ dynamical matter field reduces this to the set of confined point particles $(1,0), ...(k/2-1, 0),  (k/2,0), (-k/2 +1,0) ...   (-1,0)$, which source open line defects (flux tubes of integral electric flux).   Note that if the unscreened charge $q>k/2$, it will be screened to $q-k$ as this has a lower energy cost.


This bulk spectrum -- with $k$ distinct types of line defects --qualitatively reproduces the bulk spectrum of certain confined Abelian Walker Wang models (those based on $U(1)_{k}$ Chern-Simons theory, with $k$ even).\footnote{For $k\geq 4$, there is a quantitative difference, which is that in the gauge theory not all excitations have the same energy cost per unit length, whereas in the Walker Wang model they do.}     We will present a more detailed correspondence between the field theory presented here and these lattice models in Sect. \ref{ss:bFbbforaxioned}. 

\subsection{The obliquely confined phase for $k$  odd} \label{KoddSec}

For $k$ odd, $(-k,1)$ is a fermion, and the object that condenses as $g^2 \rightarrow \infty$ is the boson $(-2k,2)$, which has no electric charge, and flux $4 \pi$.  As a consequence, the bulk spectrum admits a deconfined fermion, analogous to the BdG quasi-particle in a superconductor, and has $\mathbb{Z}_2^f$ topological order.  
 Indeed, in Sec. \ref{sec:DiscretebF+bb}, we will show that for $k=1$ the lattice model associated with this phase is precisely the ``fermionic Toric code" argued to describe the superconductor at long wavelengths in Ref. \cite{Hansson04}  However, there is an important conceptual difference between the two: in the superconductor, the fundamental charge-carrying object is a fermion.  In the models we discuss the charge carriers are {\it bosonic}; fermions are emergent, arising because for $\theta = 2 \pi$ the neutral monopole is a composite object.

To make the comparison with the superconductor more precise, we begin by considering the case $k=1$ (i.e. $\theta = 2 \pi$), $p=2$. In the superconductor, at long wavelengths there are point-like fermionic excitations (the  BdG quasiparticles) of charge $1$ (modulo $2$), and vortex lines which carry magnetic flux in integer multiples of $\pi$ due to the Meissner effect. The BdG quasiparticle acquires a Berry phase of $\pi$ when it encircles a vortex line carrying an odd-integer multiple of $\pi$ flux.  On the lattice, where flux is defined only modulo $2 \pi$, the spectrum consists of one deconfined fermion and a single species of line defect, and is captured by a lattice model known as the fermionic Toric code (FTC)\cite{Hansson04}.  

In the limit $g^2\rightarrow \infty$ for $k=1, p=2$ in Eq. \ref{eq:effaction},  the condensate is composed of objects carrying twice the fundamental magnetic charge.  Electric flux is therefore confined to tubes carrying integer multiples of $1/2$ electric flux.  (This is the electromagnetic dual of flux quantization in a superconductor, and stems from the requirement that the Berry phase of the condensed particle with magnetic charge $2$ with the flux tube must be trivial).  The neutral monopole $(-1,1)$ is a (deconfined) fermionic excitation, which has  a Berry phase of $\pi$ if it encircles a flux tube carrying half-integral electric flux. 

Unlike magnetic flux, which is a continuous variable, it is not obvious that lines of half-integral electric flux exist at all.   In a truly compact (i.e. lattice) gauge theory where $A \equiv A+ 2\pi$, the variable conjugate to $A$ is quantized.  For $\theta =0$, this implies that half-integer electric fluxes do not exist in the quantum theory.  However, for general $\theta$ we have
\be \label{Quant1}
\left[ A_i, E_i  + \frac{\theta}{4 \pi^2} B_i \right] = i  \punc{,}
\ee 

so that it is in fact $\mathcal{E} = E  + \frac{\theta}{4 \pi^2} B$ that is quantized to be an integer.  Since in the (obliquely) confined phase electric flux is quantized in multiples of $1/2$  by the dual Meissner effect,  this means that the possible line defects carry either integer electric flux (and, for $k$ an integer, no magnetic flux modulo $2\pi$) or electric flux $(n + 1/2)$ and magnetic flux $\pi$ (mod $2\pi$).

One might expect that this would lead to a spectrum with a variety of line defects, carrying electric flux $(n + 1/2), n\in \mathbb{Z}$.  However, the monopole $(0,1)$ also carries electric charge $1$, and must be dynamical as it is a  fundamental excitation of the gauge field.  Thus even with $p=2$, screening is expected to eliminate integer electric fluxes from the long-wavelength theory, leaving only ``dyonic vortex loops" which carry $1/2,\pi$ electric and magnetic fluxes respectively. Similarly, at long distances a pair of $(1,0)$ (charge $1$) bosons will be screened by dynamical monopoles to create a pair of neutral $(-1,1)$ fermions.
Hence the obliquely confined phase for $k=1$ contains only the deconfined $(-1,1)$ fermion and the dyonic vortex loop.  As noted above, this is exactly the spectrum of the fermionic Toric code, and the mutual Berry phase between the neutral fermion and the dyonic vortex loop lead to $\mathbb{Z}_2^f$ topological order.  

It is not difficult to extend this result to general odd $k$.  Since the monopole carries charge $k$,
the minimal electrically neutral excitation $(-k,1)$ is a bound state of a monopole with an odd number of bosonic charges.  This object is a fermion and cannot condense; instead at large $g^2$  a condensate of $(-2k,2)$ will form.  The resulting spectrum contains a deconfined fermion $(-k, 1)$ and the dyonic vortex loop (carrying electric flux $1/2$ and magnetic flux $\pi/k$).  The charged $(0,1)$ monopole screens electric flux in multiples of $k$, so that now (if we take $p= 2k$) there are also non-dynamical confined test charges that source open lines of integer electric flux.  Screening by the charge $p$ dynamical matter fields will give these electric charges of  $( \pm 1,0),  ... ( \pm (k-1),0)$.     For $q > (k-1)/2$, these will be further screened by the $(0,1)$ monopole to create a composite object of charge $q-k$ bound to a neutral fermion.  
It turns out that this gives exactly the low-energy spectrum of the $U(1)_k\times \mathbb{Z}_2^f$ Walker-Wang model\cite{WWUs}, which has a $\mathbb{Z}_k \times \mathbb{Z}_2$ group structure.  

As an example, let us study the case $k=3, p=6$.  The non-dynamical charges are $( 1,0), ( 2,0), (3,0), (4,0) \equiv (-2,0)$ and $( 5,0) \equiv (-1,0)$. (Here the equivalence indicates screening by the dynamical charge $6$ boson, which will result in the configuration of minimum $E^2$ at long distances).   However, screening by the $(0,1)$ monopole (which has charge $3$) will convert these into 
\be
(\pm 1,0) \rightarrow (\pm 1,0) \ , \ \ \ (\pm 2,0) \rightarrow ( \pm 2, \mp 1) \ , \ \ \ (3,0) \rightarrow (3, -1)
\ee
  These combine according to the anticipated $\mathbb{Z}_k \times \mathbb{Z}_2$ group structure.  Note also that the electric charge of all of these defects is $\pm 1$, meaning that the energy cost of the electric flux line that connects them is the same in all cases.  Table \ref{OddTable} gives a few other examples.  Note that for $k>3$ not all confined point particles source the same amount of electric flux after screening; hence (unlike in the Walker-Wang model) in the gauge theory some test charges will be more tightly confined than others.  

\begin{widetext}
\begin{table}[h!]
\begin{center}
\begin{tabular}{|c|c|c|c|c|c|}
\hline
$k$ & fermion & Pure charges & Charge-fermion composites & Electric charge of composites  \\
\hline
3 & $(-3,1) \equiv (3,-1)$ & $(-1,0),(1,0)$ & $(-2,1), (2,-1)$ & $ q= \pm 1$ \\
5 & $(-5,1) \equiv (5,-1)$ & $(\pm 1,0),(\pm 2,0)$ & $ \pm (-3,1),  \pm (-4,1)$ & $ q= \pm 2, \pm 1$ \\
7 & $(-7,1) \equiv (7,-1)$ & $(\pm 1,0),(\pm 2,0), (\pm 3,0)$ & $ \pm (-4,1),  \pm (-5,1), \pm (-6,1)$ & $ q= \pm 3, \pm 2, \pm 1$ \\
\hline
\end{tabular}
\caption{\label{OddTable} Spectrum of the confined phase of axion electrodynamics for $p=2k$ with $k = 3,5,7$.  This spectrum contains a deconfined fermion for every $k$, as well as charges which are confined in the bulk, but deconfined on the surface.}
\end{center}
\end{table}

\end{widetext}


	\subsection{An effective topological action for the obliquely confined phase}\label{ss:bFbbforaxioned}

%
%

Having discussed the qualitative physics of the obliquely confined phase, we will now derive an effective field theory for Eq. (\ref{eq:effaction}) in the limit $g^2 \rightarrow \infty, \mac{L}_m = 0$, with the choice $p = k$ for $k$ even, and $p= 2k$ for $k$ odd.  Our derivation is somewhat heuristic: a careful derivation would provide a precise lattice regularization of the field theory \eqnref{eq:effaction}.  We will ignore for now some of the details of the underlying lattice. \footnote{Some care must be taken when defining the $F\wedge F$ on the lattice. This subtlety will not qualitatively effect the results of this section, but the correct choice must be made in order to obtain the correct periodicity in $b$ which we require to make contact with the WW models.} Consider the partition function 
\begin{align}
\label{eq:axionaction}
Z= & \int_{0}^{2\pi}\!\!dA\!\!\!\sum_{\{s_{\mu \nu},J_\mu \}}\!\!\exp\!\!\left[-\int \frac{1}{4g^{2}}\Gamma_{\mu \nu}^2+\frac{ik}{4\pi}\Gamma \wedge \Gamma
 +i p A_\mu J^{\mu}\right] 
\end{align}
associated with the effective action \eqnref{eq:effaction}. Since at every point we must sum over all possible Dirac strings (i.e. all integer $s_{\mu \nu}$),  the partition function is invariant under $F_{\mu \nu}\rightarrow F_{\mu \nu}+2\pi s_{\mu \nu}$. This periodicity in $F$ allows us
to re-write the partition function as a (discrete) Fourier series:

\be 
Z=  \int_{0}^{2\pi}dA\sum_{b_{\mu \nu},J_{\mu} \in\mathbb{Z}}e^{i \int \left (b \wedge F- p A_\mu J^\mu \right ) 
 }e^{-\widetilde{S}[b]} 
\ee 
where
\begin{align*}
e^{-\widetilde{S}[b]}= & \int_{-\infty}^{\infty}dQ e^{\int \left( -i b \wedge Q+ \frac{1}{4g^{2}}Q_{\mu \nu}^2+\frac{ik}{4\pi}Q \wedge Q\right)} \\
 = & \text{exp} \left[-  \int \left(  -\frac{i\pi}{kC}b \wedge b+\frac{2\pi^{2}}{k^{2}g^{2}C}b_{\mu \nu}^2   \right) \right ]
\end{align*}
(Such discrete actions are most naturally defined by taking our gauge theory to live on a 4D lattice, as is done for example in Ref. \onlinecite{Cardy82a}.  Here for simplicity we omit the lattice for the time being, but will re-introduce it later).  

Eq. (\ref{eq:axionaction}) is therefore equivalent to a theory with the Lagrangian density
\begin{equation}
\mac{L} = -\frac{i\pi}{kC}b\wedge b- ib \wedge F+i p A_\mu J^\mu + i \Sigma_{\mu \nu} b^{\mu \nu}  +\frac{2 k^{2}g^{2}C}{\pi^{2}}\Sigma_{\mu \nu}^2\punc{,}\label{eq:bFbbJ} 
\end{equation}
where $C=1+(\frac{2\pi}{g^{2}k})^{2}$,  $b_{\mu \nu} \in\mathbb{{Z}}$ and $F_{\mu \nu}=\partial_\mu A_\nu - \partial_\nu A_{\mu} $ with $A\in [0,2\pi )$.  
Here we have performed a (discrete) Hubbard-Stratonovich transformation to replace the $b^2_{\mu \nu}$ term with a coupling between $b_{\mu \nu}$ and a vortex field $\Sigma_{\mu \nu}$. For later reference, we note that in the Hamiltonian formulation of this theory, $\frac{1}{2}\epsilon_{i j k}\hat{b}_{i k}$ will be canonically conjugate to $\hat{A}_i$, and so (as far as the derivation above is valid) it can be identified with $-\hat{\mathcal{E}_i}$ in \eqnref{eq:Eflux}. Since we are including dynamical matter, the final step is to integrate out the matter fields.  In our model this is straightforward: summing over $J_\mu$ quantizes $A$ (and consequently $F$)  in units of $2\pi/p$.  (Adding a mass term for the matter field softens this quantization but does not qualitatively affect our discussion).  

The final result is that deep in the obliquely confined phase, in the presence of dynamical matter of charge $p$, the model (\ref{eq:effaction}) is captured by the topological field theory
\begin{equation}
S_{\text{top}} =  - i \int \left[ \frac{\pi}{k}b\wedge b + b \wedge F + j^{\mu} A_\mu +  \frac{\Sigma^{\mu \nu}}{2} b_{\mu \nu}  \right ]
\label{eq:bFbbJ2}
\end{equation}
where $A\in\frac{2\pi}{p} \{0,1,\ldots,p-1\}, \ b \in \mathbb{Z}$, together with an appropriate action for the source fields. 
Here, in addition to the vortex field tensor $\Sigma_{\mu \nu}$, we have added a (non-dynamical)  
 a matter current $j^\mu$ whose charge is defined modulo $p$.  Due to the discreteness of the fields $A$ and $b$, $\Sigma_{\mu \nu}$ takes on values in $\frac{2\pi}{l} \{0,1,\ldots,l-1\}$ where $l$ is the lowest common multiple of $k$ and $p$. 
 
 The discreteness of the various fields in \eqref{eq:bFbbJ2} implies that (at least naively) the action is periodic under $b \rightarrow b+l$, where $l=\lcm(p,k)$. (On the lattice this periodicity is somewhat subtle; see Appendices \ref{s:quantisegenk}, \ref{sec:kevenpk}). Thus $b$ is effectively a $\mathbb{Z}_{l}$ variable.  
 
Since our sources are static, we will restrict our attention to cases where only $\Sigma_{0i}, j_0$ are non-vanishing.  In this case an appropriate action for the source fields is:
\be
S_{\text{Source}} = \epsilon_p\int \mathcal{F}(e^{i \Sigma_{0i}}) + \epsilon_V  \int \mathcal{G} ( e^{i2\pi j_0/p})\punc{,}
\ee
where $\epsilon_P$ is a line tension, and $\epsilon_V$ is a mass. Here $\mathcal{F},\mathcal{G}$ are non-negative functions that vanish only if  $\Sigma_{0i}=0,j_{0} =0$ modulo $2\pi,p$ respectively. It is not difficult to see that with the correct choice of $p$ and $k$ this captures the phenomenology of the obliquely confined phases described in the previous section.

\subsection{Walker-Wang models} \label{WWSec}

In the remainder of this work, we will discuss the connection between the field theory (\ref{eq:effaction}) and the Walker-Wang topological lattice models.  Before doing this, we review some key features of these models.
For our purposes, a technical description of the general form of the Hamiltonians in question  is not necessary; instead we discuss their main phenomenological features.  In Sec. \ref{FTCSec} we will describe one Walker-Wang Hamiltonian in detail, to show how it is related to the topological field theory (\ref{eq:bFbbJ2}) in the case $k=1, p=2$.  Readers interested in the general form of the Hamiltonians are referred to  Refs. \onlinecite{Walker12,vonkeyserlingk13a}.  

Schematically, the Walker-Wang models are constructed by first choosing a 2D anyon model (a pre-modular tensor category, more generally).  
Here we will focus on the familiar case where this anyon model describes a quantum Hall state.  For example, we could consider the bosonic Laughlin state at $\nu = 1/2$.  In this case, the anyon model contains one non-trivial particle type, which we associate with the charge $1/2$ quasi-particle.  These quasi-particles are semions; two semions combine to give a trivial boson (which we call $0$) that has no interesting statistical interactions, which topologically is indistinguishable from the vacuum.  Based on this set of rules about how the anyons combine and braid with each other, an exactly solvable lattice Hamiltonian can be constructed\cite{Walker12}, which has the following properties:

\begin{enumerate}
\item {\it Confinement in the bulk}:
If our anyon model describes an allowed quantum Hall state of bosons (i.e. if the $S$ matrix of the corresponding rational CFT is modular), then there are no deconfined excitations in the 3D bulk.  
\item {\it Topological surface states}:
If the 3D lattice has a boundary, then (at least for some boundary conditions) all anyons in the underlying anyon model can appear as deconfined excitations on the boundary\footnote{There is a solvable boundary condition for which the spectrum can be calculated exactly.}.
\newcounter{enumTemp}
    \setcounter{enumTemp}{\theenumi}
\end{enumerate}

In carrying out this construction, we must be somewhat careful about what we mean by an anyon model.  If we choose a fermionic quantum Hall state, such as the $ \nu = 1/3$ Laughlin state, then the combination of three $1/3$ quasiparticles gives a fermion that braids trivially with all other anyons.  Unlike the boson of the $\nu = 1/2$ Laughlin state, however, a fermion cannot be ``trivial" topologically, and must be included in our anyon model.  (We emphasize that the fundamental charge-carrying excitations in our models are always bosons; for fermionic matter this difficulty does not arise).  Mathematically, this means that we must take two copies of our anyon model: the original, and one including the product of each anyon with this trivial fermion.  The example that we will discuss in detail here is the case $\nu=1$, where there are no anyons at all; in this case our anyon model consists of the trivial particle $0$ and a fermion.  We will refer to this anyon model as $\mathbb{Z}_{2}^{f}$.  The corresponding Walker-Wang model is an example of the following general fact\cite{vonkeyserlingk13a}:

  \begin{enumerate}
    \setcounter{enumi}{\theenumTemp}
 \item {\it Deconfined fermions in the bulk}:
If our anyon model describes a quantum Hall state of fermions (in which case there will be a fermion that braids trivially with all other anyons, and the $S$ matrix is not modular), the Walker-Wang model describes a system with a deconfined fermion in the bulk.
\end{enumerate}

Evidently, these facts suggest that a Walker-Wang model constructed from a bosonic Laughlin state describes the obliquely confined phase of (\ref{eq:effaction}) with $k$ an even integer, while the model constructed from a fermionic Laughlin state (with the ``trivial" fermion included in the set of anyons) describes the odd $k$ case.  We will make this connection more apparent in the next section, and discuss its implications in Sec. \ref{conclusion}.

\section{Discrete $bF+bb$ theories and Walker-Wang models }\label{sec:DiscretebF+bb}

Thus far, we have discussed the gapped phases of Eq. (\ref{eq:effaction}), and derived a topological action that describes the physics deep in these phases.  We will now show that $S_{\text{top}}$ is also the appropriate field theory for the Walker-Wang models discussed in Sec. \ref{WWSec}, affirming previous conjectures\cite{vonkeyserlingk13a,Kapustin13a,Kapustin13b,Walker12}.   In this section we will carry out the derivation in detail for the case $k=1, p=2$ which corresponds to the bosonic topological insulator; Appendix \ref{s:quantisegenk} explains how to generalize this result to other Walker-Wang models for integer $k$ and $p=2k$.  (As discussed in \appref{sec:kevenpk}, we are unable to derive the Hamiltonian for even $p=k$ due to subtleties involved in defining the field theory on the lattice).

Our approach is as follows.  To make a direct connection between $S_{\text{top}}$ and the lattice Hamiltonians described by Ref. \onlinecite{Walker12}, we must put our field theory on a lattice;  in \secref{ss:bFbbtoWW:Fieldtheory} we describe our conventions for doing this.  We next ask what 3D Hamiltonian gives this 4D partition function, by essentially reversing the usual derivation of the path integral in quantum mechanics.  In \secref{s:quantiseoddk} we show that the resulting Hamiltonian (for $p=2, k=1$) corresponds to the 
Walker-Wang model based on $\mathbb{Z}_{2}^{f}$.  



\subsection{Topological gauge theory on the lattice} 
\label{ss:bFbbtoWW:Fieldtheory} 

\begin{figure}
\includegraphics[width=0.95\columnwidth]{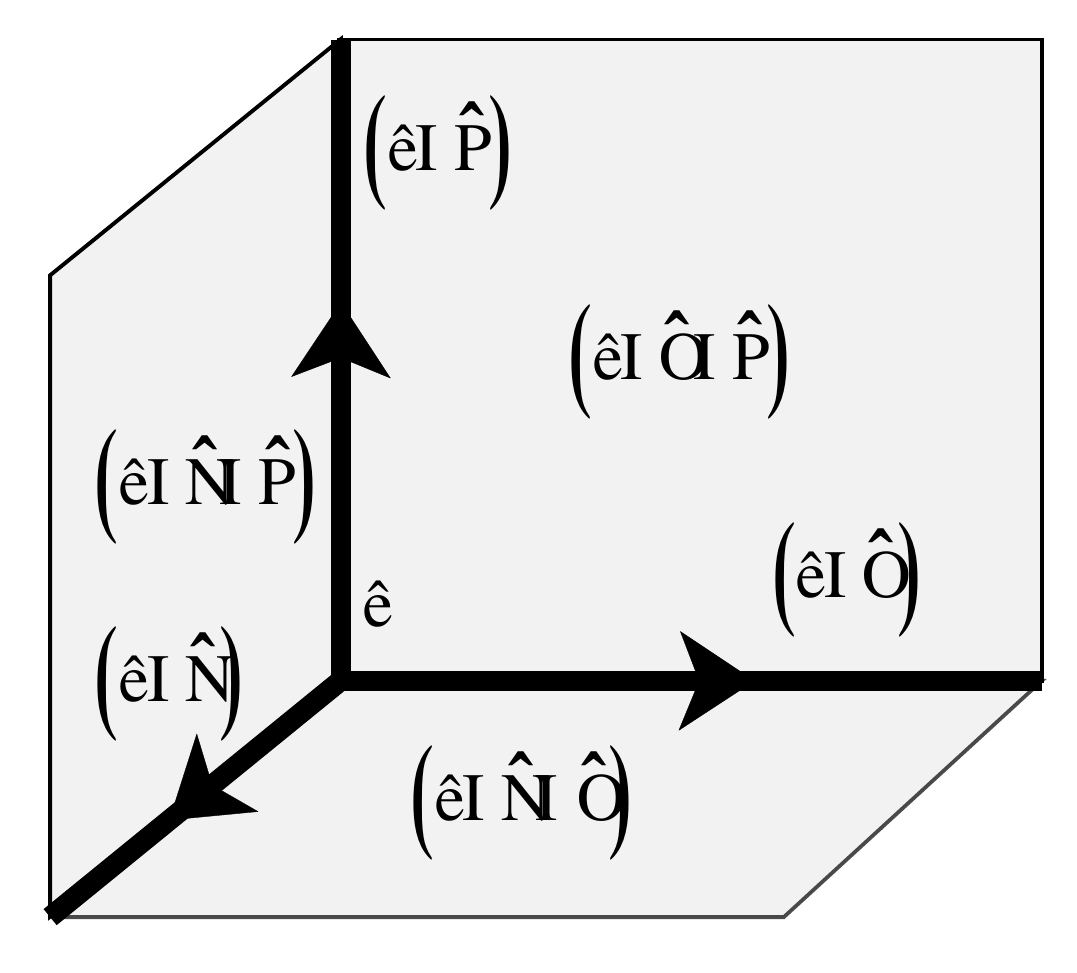}
\caption{This figure illustrates the lattice notation used
in this work, where $r=\left(r^{0},r^{1},r^{2},r^{3}\right)$ is a
vertex of the 4D lattice, while $\left(r,\hat{\mu}\right)$ denotes
an edge, and $\left(r,\hat{\mu},\hat{\nu}\right)$ a plaquette. }
\label{fig:latticenotation}
\end{figure}

We will put our 4D (space + Euclidean time) field theory  on the 4D hyper-cubic lattice $L$.  
 It is convenient to label the vertices by their co-ordinate $r=\left(r^{0},r^{1},r^{2},r^{3}\right)$, and to let $\left(r,\hat{\mu}\right)$ denote the edge connecting $r$ to $r+\hat{\mu}$. Let $\left(r,\hat{\mu},\hat{\nu}\right)$ (with $\mu\neq\nu$) denote
the plaquette bounded by $(r,\hat{\mu})$ and $\left(r,\hat{\nu}\right)$ (see Fig.~\ref{fig:latticenotation}). 

On each edge $\left(r,\hat{\mu}\right)$ of $L$, there is an associated gauge field $A_{\mu}(r)$. 
The lattice derivative $\partial_{\alpha}$ is defined by
\begin{equation}
\partial_{\alpha}T_{\mu\nu\ldots}\left(r\right)=T_{\mu\nu\ldots}\left(r+\hat{\alpha}\right)-T_{\mu\nu\ldots}\left(r\right) \ \  \hat{\alpha}=\hat{0},\hat{1},\hat{2},\hat{3}
\end{equation}
 where $T$ is an arbitrary vector or tensor field.  
 The value of the electromagnetic field strength on plaquette $\left(r,\hat{\mu},\hat{\nu}\right)$ is then given by $F_{\mu\nu}(r)\equiv\partial_{\mu}A_{\nu}(r)-\partial_{\nu}A_{\mu}(r)$, or more compactly $F\equiv dA$.

In our topological field theory there is another tensor $b_{\mu\nu}=-b_{\nu\mu}$.  This field is most naturally viewed
as living on the plaquettes of the dual lattice $L^{\star}$. The vertices of the dual lattice are formed by translating vertices of $L$ by vector $\frac{1}{2} \hat{n}$, where 
\be \hat{n}=\hat{0}+\hat{1}+\hat{2}+\hat{3} \ .
\ee
Given a vertex $r^\star$ on the dual lattice, $b_{\mu\nu}(r^\star)$ is the value of $b$ on the $(r^\star , \hat{\mu}, \hat{\nu})$ dual lattice plaquette. A plaquette $\left(r,\hat{\rho},\hat{\sigma}\right)$ on the original lattice intersects precisely one plaquette on the dual lattice lying in the $\mu\nu$ plane, where $\epsilon_{\mu\nu\rho\sigma}\neq 0$. We call this dual plaquette $(r^\star(\mu,\nu),\hat{\mu},\hat{\nu} )$ where $r^\star( \mu,\nu) = r+ \frac{1}{2}\hat{n} - \hat{\mu}-\hat{\nu}$ is a vertex in $L^*$.  This association between plaquettes and dual plaquettes is used to construct $bF$ type terms, which in this notation take the form
\be
\sum_{r} \frac{1}{4}\epsilon_{\mu\nu \rho\sigma} b_{\mu\nu}(r^\star( \mu,\nu)) F_{\rho\sigma}(r)
\ee

In  the field theory (\ref{eq:bFbbJ2}), we also have expressions of the form $b \wedge b\equiv \epsilon_{\mu\nu\rho\sigma}b_{\mu\nu}b_{\rho\sigma}$. These
involve terms like $\epsilon_{ijk} b_{ij}b_{0k}$ where $b_{0k}$ lies on time-like
plaquettes, while  $b_{i j}$ lies on space-like plaquettes. So, which
time-like plaquettes should be paired with which space-like plaquettes? It is 
natural to pair fields on plaquettes that share a vertex, but this still leaves four choices.  
Here we will use the convention:


\begin{equation}\label{eq:bblattice}
b  \wedge b\leftrightarrow\!\!\sum_{\mu,\nu,\rho,\sigma}\epsilon_{\mu\nu\rho\sigma}\frac{1}{4} b_{\mu\nu}\left(r^\star \right) b_{\rho\sigma}(r^\star- \hat{\rho}- \hat{\sigma})  \punc{.}
\end{equation}

 This prescription coincides with usual definition of $\epsilon_{\mu\nu\rho\sigma}b_{\mu\nu}b_{\rho\sigma}$
in the naive continuum limit, and defines a symmetric quadratic
form for the $b$'s.  From this point onwards, we will make a simplifying abuse of notation, and instead denote  $b_{\mu\nu}(r^\star( \mu,\nu))$ by $b_{\mu\nu}(r)$ after which, for the record, the $b  \wedge F$ and $b \wedge b$ terms become

\begin{align}
b \wedge F &\rightarrow  \frac{1}{4}\epsilon_{\mu\nu\rho\sigma} b_{\mu\nu}(r) F_{\rho\sigma}(r)\n
b \wedge b &\rightarrow \! \epsilon_{0ijk}b_{ij}\!\left(r  \right)\!\! \left[\frac{b_{0k}(r+ \hat{0}+ \hat{k})+b_{0k}(r-\hat{n} + \hat{0}+ \hat{k})}{2}\!\!\right]\punc{.}
\end{align}

%

\subsection{Hamiltonian form of $bF+bb$ theory for $k=1,p=2$}\label{s:quantiseoddk}

In this subsection we identify the Hamiltonian describing the obliquely confined bosonic topological insulator starting from the $bF+bb$ field theory (\eqnref{eq:bFbbJ2}) with $p=2,k=1$. A more general derivation for $k\in \mathbb{Z}$ runs along similar lines, and is given in App.~\ref{s:quantisegenk} and \ref{sec:kevenpk}.  Our method is to write down the partition function, express it as a Trotter decomposition, and from this deduce the effective Hamiltonian.

 With $p=2, k=1$, the $bF+bb$ Lagrangian density of \eqnref{eq:bFbbJ2} is

\begin{align}
\mac{L} &= -ib \wedge F- i\pi b \wedge b - i j^{0} A_{0} -\Sigma^{0i} b_{0i}  \nonumber\n
&   + \epsilon_V \mathcal{G}( e^{i\pi j^{0}}) + \epsilon_{P}  \mathcal{F}(e^{i\Sigma^{0i}})  \punc{,}
 \label{eq:kis1}
\end{align}

where $F
=d A$, $A\in \{0, \pi \} $, and
$b_{0i},\frac{1}{2}\epsilon_{ijk}b_{jk}\in\left\{ 0,1\right\} $.  Here we take these sources to be non-dynamical, and restrict our attention to the case where the spatial components $j_i, \Sigma_{ij}$ of both sources vanish.\footnote{Including spacelike sources will introduce operators creating the corresponding excitations in the Hamiltonian.} Both $j^{0}$ and $\Sigma^{0i}/\pi$ are integers modulo $2$, representing static charge and line-defect sources. As they do not explicitly depend on $A,b$, we will drop $\mathcal{F},\mathcal{G}$ for now and re-introduce them at a later stage.

The first step is to ``integrate out" (i.e. sum over) $A_0$, which imposes the `Gauss' law constraint' $\epsilon_{i j k} \partial_i b_{j k} =J_0\mod 2$.  Having eliminated the non-dynamical field $A_0$ at the expense of imposing a constraint, 
the partition function can now be expressed:

\begin{align}
\sum_{\{b\}}'\sum_{\{\tilde{A}\}}\exp\left[i\sum_{r}( b \wedge d\tilde{A}+  \pi b\wedge b+\Sigma^{0i} b_{0i} )\right] \punc{,}
\end{align}
where $\sum'$ is a sum over the constrained $b$ fields, and where $\tilde{A}_{\mu}$ contains just the space-like components of $A_\mu$. 

At this point, it is convenient to introduce a change of variable names. Define Ising fields $\sigma^z,\sigma^x=\pm1$ living on the space-like edges of the hyper-cubic lattice. (These fields are classical, but the reason for their names will become clear shortly!)  Set 
\be
b_{i j}(r)= \epsilon_{ i j k } (1-\sigma^{z}_{r,k})/2 \ , \ \   A_{i}(r)= \pi(1-\sigma^{x}_{r,i})/2
\ee
 for the space-like components of $b$ and $A$. In terms of the new variables, the Lagrangian density is

\begin{align}
& \sum^3_{j=1}\frac{i \pi}{4} (1-\sigma^z_{r,j})\partial_0 (1-\sigma^x_{r,j})    - \sum^3_{k=1} i\Sigma^{0k}(r) b_{0k}(r) \n
&-\sum_{k=1}^{3} \frac{i \pi}{4}b_{0k}(r) \left[  (1-\sigma^z_{r- \hat{0}- \hat{k},k})\!+\!(1-\sigma^z_{r+\hat{n}-\hat{0}-\hat{k},k}) \right] 
\n
&- \sum_{i,j, k :  \epsilon_{i j k } =1}\frac{i \pi}{4} b_{0 k}(r)  \db{1-\!\!\prod_{e\in \partial (r,\hat{i},\hat{j}) }\!\sigma^{x}_{e} }\punc{,}
\end{align}

where $\partial (r,\hat{i},\hat{j})$ denotes the edges in the boundary of plaquette $ (r,\hat{i},\hat{j})$. The Gauss constraint can be rewritten $\mathcal{Q}_r \equiv \prod_{e\in s(r)} \sigma^z_e = (-1)^{J_0(r)}$, where $s(r)$ is the set of six space-like edges attached to vertex $r$. 

The second step is to sum out the (unconstrained) time-like components of $b$ (i.e. $b_{0k}$).  An integration by parts then gives a partition function 

\begin{align}
 & \sum'_{\{\sigma^z\}}  \sum_{\{\sigma^x\}} \prod_{ (r,\hat{i},\hat{j})} \left[1+e^{i\pi \Sigma^{0 k}(r)} \prod_{e\in \partial (r,\hat{i},\hat{j}) } \! \!\!\!\sigma^{x}_{e} \times \sigma^{z}_{r+\hat{i}+\hat{j},k}\sigma^{z}_{r-\hat{k}-\hat{0},k}\right]  \nonumber \\
 &  \times \exp\left[\sum_{r,i} \frac{i\pi}{4}\left(\sigma^{z}_{r,i}-\sigma^{z}_{r-\hat{0},i}\right)(1-\sigma^{x}_{r,i})    \right]\punc{.}\label{eq:oddkpartitionnosource}
\end{align}
Here the first product is over the space-like plaquettes present in the $ij=12,23,31$ planes at every time slice; within this product, the direction $\hat{k}=\hat{1},\hat{2},\hat{3}$ is uniquely specified by requiring $\epsilon_{ i j k}=1$. Since $b_{0k}$ is also a non-dynamical field, this sum imposes a second constraint: the partition function vanishes unless
\be
\mathcal{B}_{(r,\hat{i},\hat{j})} \equiv\prod_{e\in \partial (r,\hat{i},\hat{j}) } \! \!\!\!\sigma^{x}_{e} \times \sigma^{z}_{r+\hat{i}+\hat{j},k}\sigma^{z}_{r-\hat{k}-\hat{0},k}= e^{-i\pi \Sigma^{0k}(r)} \punc{.}
\ee

Using this constraint and the Gauss condition, and re-introducing the $\epsilon_V,\epsilon_P$ terms in \eqnref{eq:kis1} allows us to remove all explicit dependence on $\Sigma_{0i},j_0$

 \begin{align}
 & \sum'_{\{\sigma^z\}}  \sum_{\{\sigma^x\}} \exp\ds{\epsilon_v \sum_r \mathcal{G}\db{\mathcal{Q}_r} + \epsilon_p \!\!\!\sum_{(r,\hat{i},\hat{j})} \mathcal{F}\db{\mathcal{B}_{(r,\hat{i},\hat{j})}}  } 
 \nonumber \\
 &  \times \exp\left[\sum_{r,i} \frac{i\pi}{4}\left(\sigma^{z}_{r,i}-\sigma^{z}_{r-\hat{0},i}\right)(1-\sigma^{x}_{r,i})    \right]\punc{.}\label{eq:oddkpartition}
\end{align}
where (without loss of generality\footnote{If $\kappa=\pm 1$, then $\mathcal{F}(\kappa),\mathcal{G}(\kappa)$ can be written generally as $\alpha + \beta \delta_{\kappa,1}$}) we will take $\mathcal{F}(\kappa)=\mathcal{G}(\kappa )=2\delta_{\kappa,1}-1$.

 We now express \eqnref{eq:oddkpartition}  as $\tr e^{-\hat{H}_{\text{eff}}\tau}$
where $\hat{H}_{\text{eff}}$ is a 3D quantum Hamiltonian. The full details of this procedure are in \appref{s:quantisegenk}, but here are the key steps. Note that a spin-$1/2$ system has obvious bases $\hat{\sigma}^\alpha \ket{\sigma^\alpha} =\sigma^\alpha \ket{\sigma^\alpha} $ for $\alpha=z,x$. It is easy to verify that
\be
\left\langle \sigma'^{z}\right.\left|\sigma^{x}\right\rangle \left\langle \sigma^{x}\right|\left.\sigma^{z}\right\rangle =\exp\left[\frac{i\pi}{4}\left(\sigma'^{z}-\sigma^{z}\right)\left(1-\sigma^{x}\right)\right] \punc{.}
\ee
We can use this identity to replace the last term of the second line in \eqnref{eq:oddkpartition} by 

\be\label{eq:undotrotter}
\prod_n
\left\langle \{\sigma^{z,(n)}\}\right.\left|\{\sigma^{x,(n)}\}\right\rangle \left\langle\{ \sigma^{x,(n)}\}\right|\left.\{\sigma^{z,(n-1)}\}\right\rangle \punc{,}
\ee
where $\left|\{\sigma^{\alpha}\}\right\rangle=\otimes_e \left|\sigma^{\alpha}_e\right\rangle$, with $e$ labeling the space-like edges of the cubic lattice. The index $n$ labels the time-slice, corresponding to the $r^0$ co-ordinate in \eqnref{eq:oddkpartition}. With care, one can replace appearances of $\sigma^z,\sigma^x$ in $\mathcal{B},\mathcal{Q}$ in \eqnref{eq:oddkpartition} with operator insertions of $\hat{\sigma}^z,\hat{\sigma}^x$ between the bras and kets in \eqnref{eq:undotrotter}.  Effectively the classical variables $\sigma^{z}_{i},\sigma^{x}_{i}$ are promoted to the quantum mechanical operators $\hat{\sigma}^{z}_{i},\hat{\sigma}^{x}_{i}$, obeying the familiar Pauli-matrix commutation relations. Making these replacements, the partition function is a Trotter decomposition of a quantum mechanical problem with Hamiltonian

\begin{equation}
\hat{H}_{\text{eff}}\Delta \tau =-\epsilon_P \sum_{P} \ub{\prod_{e\in\partial P}\hat{\sigma}_{e}^{x}\times\hat{\sigma}_{O(P)}^{z}\hat{\sigma}_{U(P)}^{z}}_{\hat{B}_{P}} - \epsilon_v \sum_V \ub{\prod_{e \in s(V) } \hat{\sigma}^{z}_{e}}_{  \hat{Q}_V} \punc{,}\label{eq:FTCHamiltonian}
\end{equation}

which is  precisely the Fermionic toric code Walker-Wang
Hamiltonian\cite{Walker12,vonkeyserlingk13a,Burnell13}. Hence the discrete $bF+bb$ theory with $k=1,p=2$ (\eqnref{eq:kis1}) is captured by a known Walker-Wang model. 

Let us now review the physics of our Hamiltonian (\ref{eq:FTCHamiltonian}). Since we have eliminated the sources from the problem, the Hilbert space of this Walker-Wang model consists of a two state system $\sigma_{e}^{z}=\pm1$ on each edge $e$ of a cubic lattice. 
Henceforth, we represent configurations by coloring in edges with $\sigma^z_e=-1$, and leaving $\sigma^z_e=1$ edges empty. $P$ labels plaquettes on the cubic lattice and $V$ the vertices.  

 The Hamiltonian consists of a sum of vertex operators and plaquette operators. The vertex operators $\hat{Q}_V$ take values $\pm1$ if there are an even/odd number of down spins on the six edges (denoted $s(V)$) attached to vertex $V$. In the field theory there is a Gauss' law constraint is $\hat{Q}_V=(-1)^{J_0(V)}$, meaning that there are an odd number of  down spins on the edges coming into vertex $V$ if and only if there is matter sitting at this vertex, and the energy cost of this configuration is that of the matter field.  In the Hamiltonian formulation we have dropped the matter fields entirely in favor of the edge variables; the two formulations are obviously equivalent.  States for which $\hat{Q}_V$ has eigenvalue $1$ can be represented as closed loops.
 
\begin{figure}
\begin{centering}
\includegraphics[width=0.9\columnwidth]{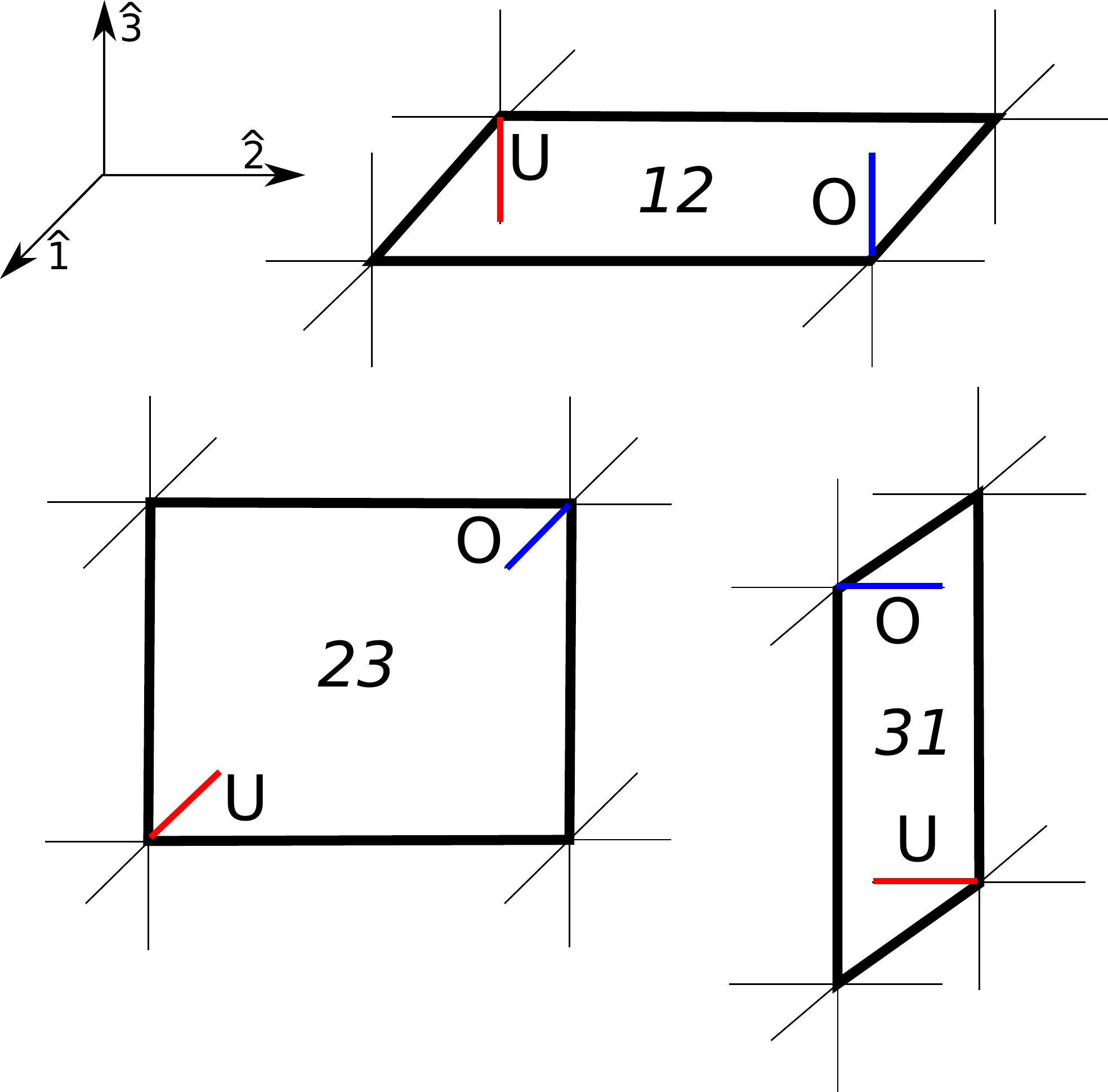}\\
 \caption{(Color online): For each of the three types of plaquettes (in the $12$,
$23$ and $31$ planes) this figure denotes two privileged edges used
to define the operator $\hat{B}_{P}$ in \eqnref{eq:FTCHamiltonian}. }
\label{fig:plaquette} 
\par\end{centering}

\centering{}
\end{figure}

 The plaquette operator is as follows. Given a plaquette $P$, we let $\partial P$ denote the four edges on the boundary of $P$. There are also two privileged edges $O(P),U(P)$ which depend on whether the plaquette lies in the $12$, $23$ or $31$-plane as shown in \figref{fig:plaquette}. $\hat{B}_{P}$  flips the spins on each of the 4 edges in $\partial P$, and multiplies by a phase $\hat{\Theta}_{P}=\hat{\sigma}_{O(P)}^{z}\hat{\sigma}_{U(P)}^{z}$ which depends on the state of the two $O(P),U(P)$ edges. It is easy to verify that $[\hat{B}_{P},\hat{Q}_V]=0$, and moreover that $[\hat{Q}_V,\hat{Q}_{V'}]=[\hat{B}_{P},\hat{B}_{P'}]=0$. This model is almost identical to the usual toric code, with the only difference being the plaquette term: The toric code has $\hat{B}_{P}^{TC}=\prod_{e\in\partial P}\hat{\sigma}_{e}^{x}$ while the FTC has $\hat{B}_{P}=\hat{B}_{P}^{TC}\times\hat{\Theta}_{P}$. In the next section we explain some of the notable properties of this FTC, and their relationship with the bosonic topological insulator.

\label{s:bFbbtoWW} 

%

\section{A case study: an obliquely confined bosonic topological insulator and the fermionic toric code.} \label{FTCSec}

The Hamiltonian derived in the previous section corresponds exactly to the ``fermionic Toric code" (or FTC). This Walker-Wang Hamiltonian is constructed from the anyon model  $\mathbb{Z}_2^{f}$ (i.e., $U(1)_1$ Chern-Simons theory, which describes a $\nu=1$ bosonic Laughlin state, together with the emergent fermion required to make it well-defined), and contains only a single species of fermion.  We now verify that this model behaves like the strongly gauged bosonic topological insulator described in Section \ref{KoddSec}.  This exploration gives readers a flavor for the models, and for how the general features of their spectra mentioned in Sec. \ref{WWSec} arise.  Then in \secref{ss:WWtoXu} we connect the ground states of the fermionic Toric code to a picture of the bosonic topological insulator ground state derived from a non-linear $\sigma$ model\cite{Xu13}. 

\subsection{Ground states}\label{ss:bFbbtoWWGS} 
\begin{figure*}
\includegraphics[width=1\linewidth]{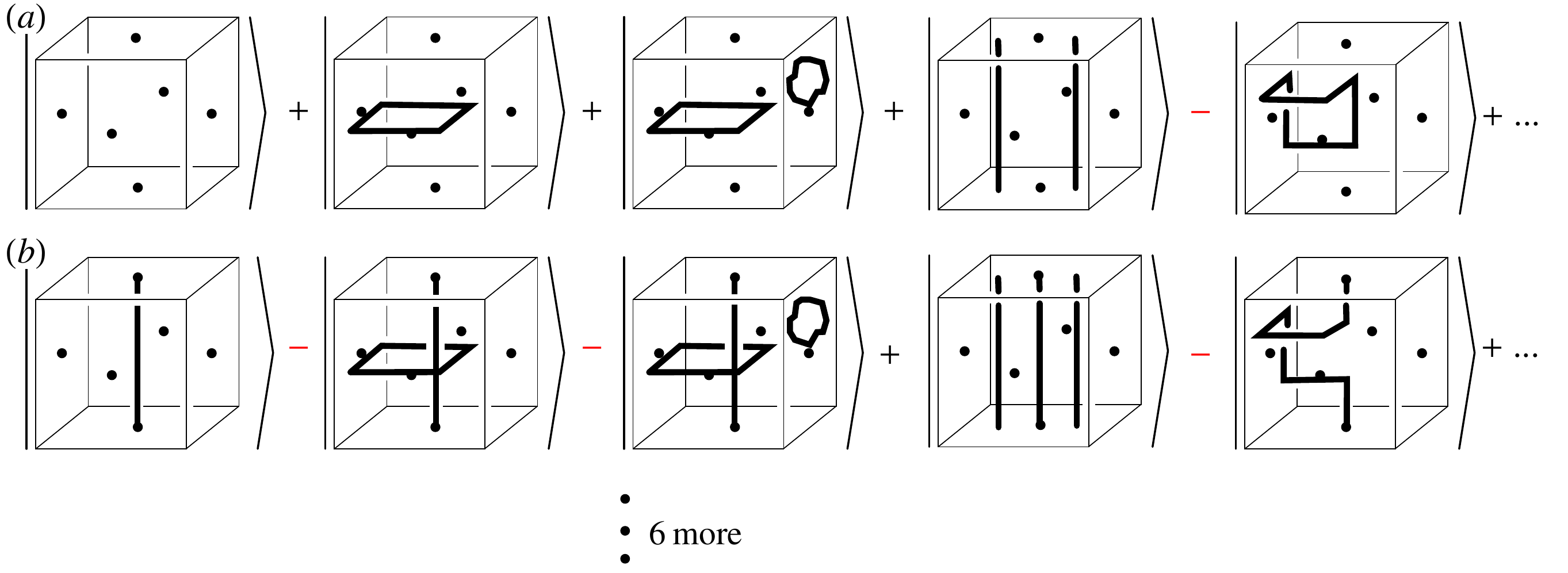}\\
\begin{centering}
\caption{(Color online): This figure shows two of the $2^3$ FTC ground states on the 3-torus, with the underlying lattice not drawn, for simplicity. The ground states are distinguished by whether an even or odd number of loops wind around the non-contractible cycles of the torus. Configurations with an odd `crossing parity' appear with $-1$ amplitude in the ground state superposition.}
\label{fig:FTCGS} 
\par\end{centering}
\end{figure*}

The ground states of \eqnref{eq:FTCHamiltonian} are defined by $\hat{Q}_V=1$
and $\hat{B}_{P}=1$ for all vertices $V$ and plaquettes $P$. The first condition implies that the ground
state is a superposition of closed loops%
\footnote{To define a notion of `loop' on the edges of a cubic lattice, one
actually needs to point-split trivalent vertices as shown in Refs.~\onlinecite{vonkeyserlingk13a,Walker12}%
}. We now show how the second condition determines the relative amplitudes
of loop configurations in the ground state superposition. As an example, the
following equation follows from the definition of $\hat{B}_{P}$,

\begin{equation}
\mathord{\includegraphics[width=\columnwidth]{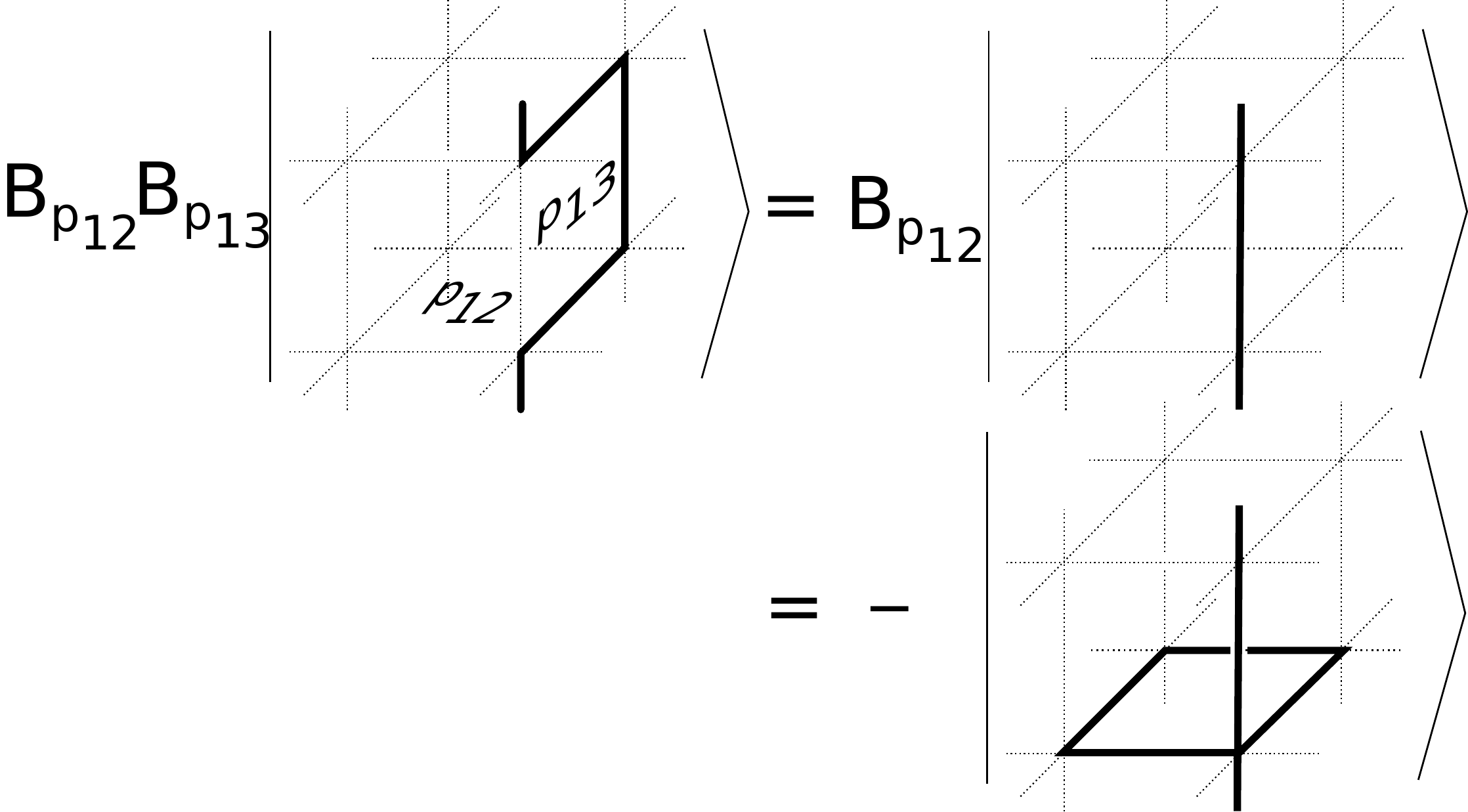}}\punc{.}\label{eq:FTCplaquetteflips}
\end{equation}

The condition $\hat{B}_{P}=1\forall P$ implies that, within the ground
state superposition, the first configuration arises with the same
amplitude as the second, while the third
arises with a $-1$ phase relative to the first two. More generally,
if two loop configurations are related by flipping a plaquette $P$,
then they have relative sign $\hat{\Theta}_{P}=\hat{\sigma}_{O(P)}^{z}\hat{\sigma}_{U(P)}^{z}$ in the ground state.
 
 Fortunately, a simple graphical mnemonic determines
the relative signs of loop configurations in the ground state. Consider a configuration of closed
loops $\{\sigma_{e}^{z}\}$. We say it has even/odd `crossing parity' if, on viewing the configuration from the $(1,1,1)^T$ direction, we see an even/odd number of over-crossings of strings. In the ground state superposition, states with odd crossing parity arise with a $-1$ relative to those with even crossing parity. For instance, in \eqnref{eq:FTCplaquetteflips}, the first configuration has no over-crossings, while the third has a single over-crossing, so these two states occur with a relative $-1$ sign in the ground state. 

In summary,  use the following
procedure to form a ground state of \eqnref{eq:FTCHamiltonian}. Start with any configuration of closed loops $\ket{\!\!\{\sigma^z\}}$, and generate
related configurations by acting on it with all
possible combinations of plaquette flips. Sum up all the generated configurations,
remembering to weight them with a $+1 / -1$ coefficient if they have even/odd crossing parity (\figref{fig:FTCGS}). There is a ground state
degeneracy on non-simply connected manifolds because not all closed loop configurations are related by plaquette flips. For example, on the 3-torus, plaquette flips
cannot change the number of loops modulo $2$ winding around the non-contractible
cycles of $\mathbb{T}^{3}$ ( \figref{fig:FTCGS}). This leads to an $2^{3}$-fold ground
state degeneracy on $\mathbb{T}^{3}$, with each ground state labelled by the three independent winding
number parities.

\subsection{Excitations}\label{ss:bFbbtoWWEx} 
Here we briefly overview the properties of the excitations --- for a more detailed treatment see Ref.~\onlinecite{vonkeyserlingk13a}. The Hamiltonian \eqnref{eq:FTCHamiltonian} has two types of excitations: pairs of vertex defects where $\hat{Q}_V=-1$, and loops of plaquettes defects (`vortex lines') where $\hat{B}_{P}=-1$. As in the standard toric code\cite{Hamma05}, vertex defects acquire a $\pi$ berry phase under encircling a line of plaquette defects.

\begin{figure}
\includegraphics[width=0.8\linewidth]{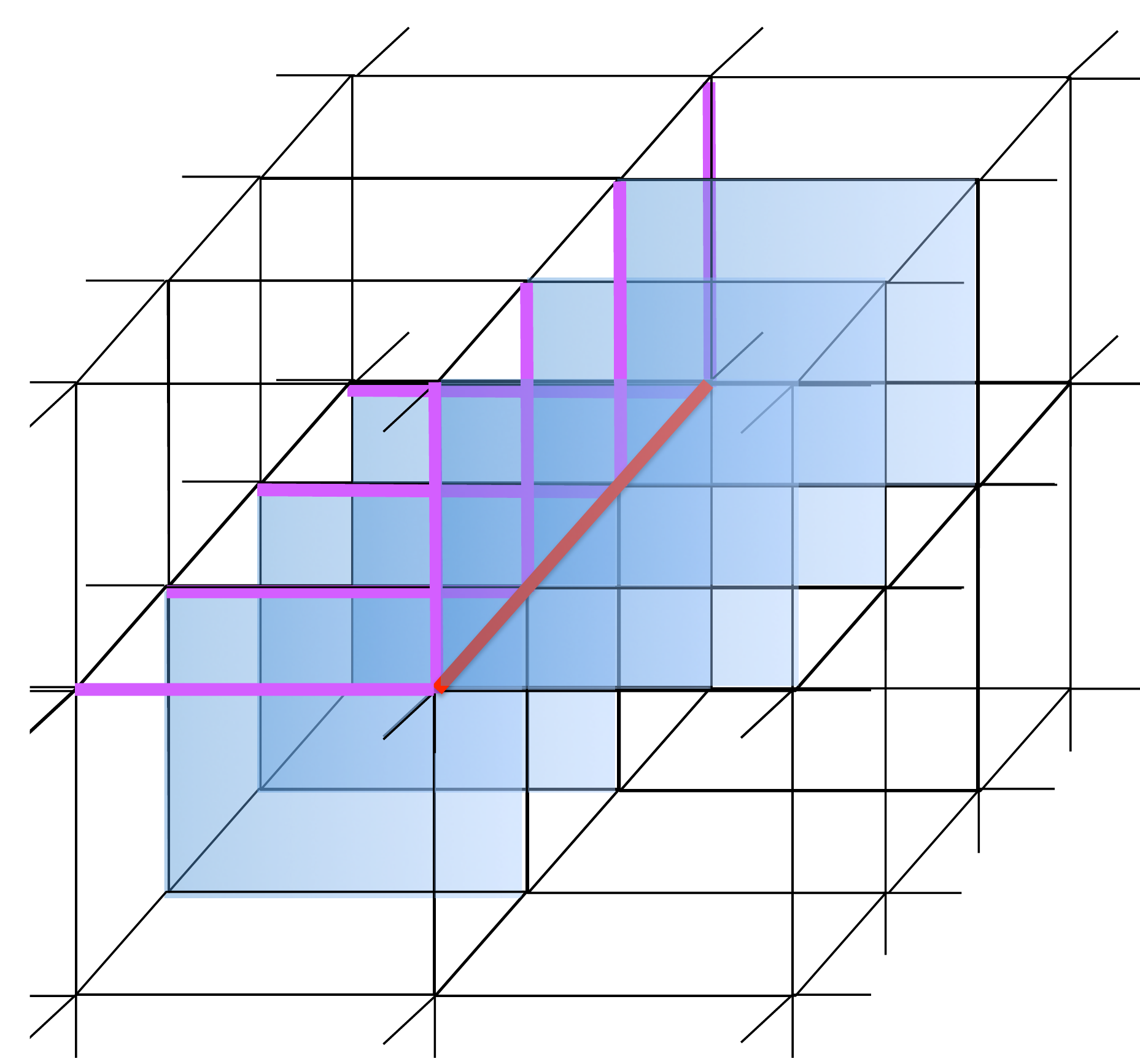}\\
\begin{centering}
\caption{(Color online):  The string operator that creates deconfined fermions in the bulk acts with $\sigma^x$ along a series of edges (shown here in red) creating a pair of vertex defects at its endpoints.  To make this operator commute with the plaquette terms in the Hamiltonian, it must also act with $\sigma^z$ on an adjoining set of ``crossed edges", shown here in purple.  This cancels the possible plaquette defects created by acting with $\sigma^x$ on the $O$ or $U$ edges of the adjoining plaquettes shaded blue.}
\label{DefectFig} 
\par\end{centering}
\end{figure}

The difference between the toric code and FTC is simply in the statistics of vertex excitations: the pairs of deconfined vertex defects are bosonic in the TC, but fermionic in the FTC. This follows from the form of the operator
that creates the deconfined fermionic excitations 
\begin{equation}
W_{V}^{f}(\maC)=\prod_{j\in\mathcal{C}}\hat{\sigma}_{j}^{x}\underbrace{\prod_{\text{crossed edges}}\hat{\sigma}_{i}^{z}}_{\hat{\Phi}(\maC)}\punc{.}\label{eq:FTCstring}
\end{equation}
The excitations lie at the endpoints of the path $\mathcal{C}$. The
first term in $W_{V}^{f}(\maC)$ is the same operator that creates deconfined bosonic excitations
in the toric code, by flipping all the edges along the path $\mathcal{C}$. The second term $\hat{\Phi}(\maC)$ is required to make this operator commute with the FTC plaquette term. Consider shifting $\mathcal{C}$ along the projection
direction $\pm(1,1,1)^T/2$ to form two paths $\mathcal{C}_{\pm}^{*}$
on the dual lattice (Fig. \ref{DefectFig}). Each shifted copy of $\mathcal{C}$ pierces a string of plaquettes on which our string operator has flipped $\sigma^z$ on either an $O$ or a $U$ edge.  The crossed edges are chosen such that $\hat{\Phi}(\maC)$ exactly cancels
the resulting plaquette defects 
without violating additional plaquettes.  
(see Ref.~\onlinecite{vonkeyserlingk13a} for details).  
One can check that $W_{V}^{f}(\maC)$ commutes with the Hamiltonian everywhere except
near its endpoints. However, the additional phases due to $\hat{\Phi}(\maC)$
also imbue the vertex defects with fermionic statistics.

\subsection{Connection between axion gauge theory and Walker-Wang degrees of freedom}\label{ss:connectiontoBTI} 
In \secref{KoddSec}, we saw that the dyon condensed $k=1,p=2$ axion theory had fermionic $(-1,1)$ deconfined excitations, and $E_i=1/2$ electric flux defects. These can be identified with the vertex defects, and lines of plaquette defects respectively in the fermionic toric code.

Let us establish a concrete dictionary between the degrees of freedom in the axion theory and those in the Walker-Wang model. First, the $\hat{\sigma}^x_i$ operator in the WW model is associated with $e^{i \hat{A}_i}$ in the $bF+bb$ theory, which is also associated with $e^{i \hat{A}_i}$ in the axion theory. Second, notice that the operator that measures whether or not an edge contains a string ($\hat{\sigma}^z_i$) is associated with $e^{\frac{i\pi}{2}\epsilon_{i j k} \hat{b}_{jk}}$ in the $bF+bb$ field theory, which is conjugate to $\hat{A}_i$ in \eqnref{eq:bFbbJ}. However, in \eqnref{eq:Eflux} and \eqnref{Quant1}, the variable
\be\label{eq:cont}
\hat{\mathcal{E}}_i  \equiv  \hat{E}_i + \frac{\theta}{4\pi^2} \hat{B}_i
\ee
is canonically conjugate to $\hat{A}_i$ in the axion field theory. Hence, to the extent that our derivation of the $bF+bb$ theory from the axion theory is valid, we can associate $\hat{\sigma}^z_i$ in the Walker-Wang model with $e^{-i\pi \hat{\mathcal{E}}_i }$ in the axion field theory. 

With the dictionary established, let us interpret the Walker-Wang plaquette operator (\eqnref{eq:FTCHamiltonian}) in the axion model language. The plaquette operator consists of two parts. The first part measures $\prod_{e\in \partial P} \hat{\sigma}^x$, which in the axion model is just the value of $e^{i \hat{B}_i}$ where $\hat{B}_i = \epsilon_{i j k }\hat{F}_{j k}/2$ is the magnetic flux through the plaquette $P$. The second part is a phase $\hat{\sigma}^z_{O(P)} \hat{\sigma}^z_{U(P)} = \pm 1$, which, in the axion model can be interpreted as total value of $e^{-i\pi \hat{\mathcal{E}}_i }$ transverse to the plaquette: $\exp\ds{-i \pi \db{\mathcal{E}_{O(P)} +\mathcal{E}_{U(P)}}} $.\footnote{There are other ways to define this total flux, but the point-splitting convention chosen in \eqnref{eq:bblattice} demands we adopt this definition.} On net, the fermionic toric code plaquette operator $\hat{B}_{P}$ corresponds to the measurement $e^{-i 2 \pi \hat{E}^{\text{PS}}_i}$, where 
\be
\hat{E}^{\text{PS}}_i \equiv   \frac{\mathcal{\hat{E}}_{O(P)} +\mathcal{\hat{E}}_{U(P)}}{2}  - \frac{\theta}{4\pi^2} \hat{B}_i 
\ee
can be interpreted as a lattice regularized or point-split (PS) analogue of \eqnref{eq:cont}. In the present case note that $\frac{\theta}{4\pi^2}=\frac{1}{2\pi}$. The ground state of the fermionic toric code thus has $\hat{B}_{P}=1$, which corresponds to $\hat{E}^{\text{PS}}_i=0  \mod 1$. The plaquette line defects, on the other hand, are associated with $\hat{B}_{P}=-1$, or electric flux $\hat{E}^{\text{PS}}_i = \frac{1}{2} \mod 1$ in the axion model. These two statements are in full agreement with the discussion in \secref{KoddSec}. 

What about the vertex defects? Within the Walker-Wang model, the operator creating these defects is explained in \eqnref{eq:FTCstring}. Using the dictionary established above, we can interpret a vertex defect creation operator as 
\be
e^{i \int_{\mathcal{C}} \hat{A}_i} \times e^{-i \pi \sum_{\text{crossed edges}} \hat{\mathcal{E}}_i }
\ee
in the axion gauge theory. The first part of this operator toggles $ \hat{\mathcal{E}}_i $ by $-1$ along the path $\mathcal{C}$. The second part creates two $-\pi$ magnetic flux tubes on plaquettes adjacent to $\mathcal{C}$ (precisely the shaded plaquettes in \figref{DefectFig}). This can be viewed as creating a point-split flux tube carrying magnetic flux $-2\pi$. The tube carries no net electric flux because, averaged over a few plaquettes, the electric field is $E_i =\mathcal{E}_{i} - \frac{\theta}{4\pi^2} B_i  =-1 + 1 =0$ modulo $1$. In the axion field  theory, the neutral monopole $(-1,1)$ has exactly the same property -- it exudes no electric flux, but carries magnetic flux $2\pi$. The fermionic statistics of this excitation follow from an argument in Ref.~\onlinecite{Goldhaber76}. 

\subsection{Connection to wave function of Bosonic topological insulators}\label{ss:WWtoXu} 
\begin{figure}
\includegraphics[width=1\linewidth]{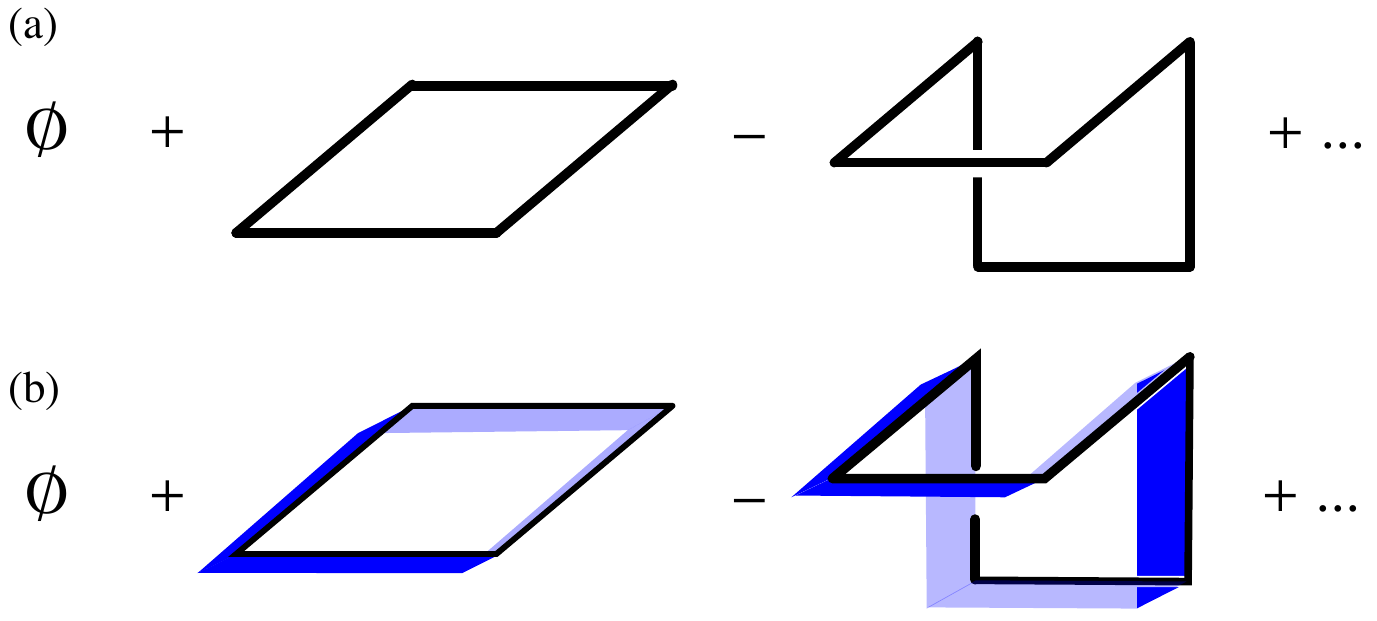}\\
\begin{centering}
\caption{(Color online): This picture shows how the loops in the Walker-Wang model ground state (a) can be made into ribbons (b) by first duplicating each loop, and then dragging the copies in the $\vec{n}$ direction. The $-1$ phases arise in the ground state precisely when these ribbons are twisted an odd number of times. In the final configuration in $b$ this twist manifests itself in the fact that one end of the ribbon links the other end precisely once, leading to phase $-1$. The ribbon has two sides, which are light/dark blue.}
\label{fig:FTCGSribbons} 
\par\end{centering}
\end{figure}

In Ref.~\onlinecite{Xu13}, the authors present the ground state of the bosonic topological insulator as a condensate of vortex ``ribbons" in which phase factors $(-1)$ appear when the ribbon is twisted an odd number of times.  In this section, we show that this picture of the ground states persists in the FTC, which describes the gapped strong-coupling phase of the same field theory. 

We briefly review the picture of Ref.~\onlinecite{Xu13}. The point of departure is to view a bosonic insulator in the phase basis $\ket{\{\hat{\phi}\}}$, using the standard number-phase representation of the Bose operators $\hat{b}_i=\sqrt{\hat{\rho}_i} e^{i \hat{\phi}_i}$. In the strongly insulating limit $\hat{\rho}$ fluctuates little, and so the conjugate phase variables $\hat{\phi}$ are maximally disordered. Indeed, we can think of the insulating state as being a condensate of $2\pi$ vortex defects in the phase variables.

The bosonic topological insulator is conveniently understood in this language. The $U(1)\rtimes \mathbb{Z}_2^T$ BTI is most easily described using two species of bosons, and therefore two species of vortex defects, which always appear in tightly bound pairs to give a $U(1)$ (rather than $U(1) \times U(1)$) symmetry.  In such a bound state, the two vortex loops can be thought of as defining two edges of a ribbon. Any given ribbon will twist by some multiple of $2\pi$ radians over its length. The ground state of the BTI differs from a trivial insulator in that it has a $-1$ phase whenever the ribbons twist by an odd multiple of $2\pi$ radians in total. 

To obtain the fermionic toric code, we gauged the BTI considered in Ref.~\onlinecite{Xu13} and set $g^2 \rightarrow \infty$, gapping the gauge field to leave residual $\mathbb{Z}_2$ theory discussed in \secref{sec:confinedBTI}.  As discussed above (see \figref{fig:FTCGSribbons}(a)), we can view the ground state of the Fermionic toric code as a superposition of closed loops, weighted by $(-1)^{\text{over-crossings}}$ where over crossings are determined by viewing the lattice from the $(1,1,1)^T$ direction. To make contact with Senthil-Xu's `ribbon' picture, imagine turning a configuration of loops in the WW ground state into a configuration of ribbons by first duplicating each loop, and then shifting the duplicate loops by $(1,1,1)^T/2$  as in \figref{fig:FTCGSribbons}(b). If a loop has an odd number of over-crossings, it turns out that the ribbon thus defined has an odd number of twists. Thus, the phase of the configuration in the ground state superposition is $(-1)^{\text{over-crossings}}=(-1)^{\text{self-twists}}$ as shown in  \figref{fig:FTCGSribbons}(a) and (b). 

To make the correspondence more concrete, note that the $\sigma^z =-1$ loops in the WW model correspond to the edges on which $b_{ i j} =1$ (mod $2$) in the field theory (\eqnref{eq:kis1}). In the ground state (where $\Sigma_{0i}=0$), however, the sum over $b_{0i}$ imposes the constraint

\be
\frac{1}{\pi}F_{i j}(x+\hat{i}+\hat{j}) =  b_{i j }(x+\hat{n}) + b_{i j }(x)  \mod 2 \punc{.}
\ee
This equation tells us that a single edge with $b_{i j}(x)=1$ is associated with a pair of $F=\pi \mod 2\pi$ magnetic flux tubes on plaquettes $(x+\hat{i}+\hat{j}, \hat{i},\hat{j})$ and $(x+\hat{i}+\hat{j}-\hat{n}, \hat{i},\hat{j})$. Hence the $\sigma^z=-1$ loops in the WW ground state are associated with pairs of $\pi$ magnetic flux tubes (i.e. pairs of line-like plaquette defects for which $\prod_{e\in \partial P} \hat{\sigma}^{x}=-1$) separated by the vector $\hat{n}$. The pair of $\pi$ flux tubes can be viewed as depicting a point split $2\pi$ flux tube. So, the ground state looks like a superposition of $2\pi$ flux tubes, bound to a $b_{ij}=1$  loop. The resulting ribbon-like objects are weighted by the phase $(-1)^{\text{self-twisting}}$, exactly as in the BTI [\onlinecite{Xu13}]. While the ribbons in [\onlinecite{Xu13}] are bound states of two vortex loops corresponding to two species of bosons, the WW ribbons are bound states of two $\pi$ magnetic flux loops, and the $b_{ij}=1$ loops.

\section{A few odds and ends} \label{RandomSec}

Before concluding, let us pause to address some questions that readers might have at this juncture.  First, one of the defining features of the Walker-Wang models that we consider (with the exception of the fermionic Toric code) is that they have excitations with anyonic statistics confined to their surfaces. How do these anyonic excitations arise in the gauge theory, at the interface between two confined phases with different (integer) values of $k$?  
Second, we will discuss time reversal symmetry, which one expects in a field theory with $\theta = 2 \pi k$, but which is naively not present in the Walker-Wang Hamiltonians beyond the fermionic toric code.

\subsection{Deconfined surface anyons at the boundary between confining phases}\label{sec:Surface}
 The correspondence that we have established between axion electrodynamics and Walker-Wang Hamiltonians suggests that the boundary between a Walker-Wang model and the vacuum is equivalent to the interface between two confined phases of axion electrodynamics: one with $\theta = 2 \pi k$ and one with $\theta =0$.  In the former case, this boundary hosts deconfined anyonic excitations for $k\geq 2$, which are otherwise confined in the bulk.  Here we discuss how this can arise from the field theoretic point of view.  

There are two questions we must address: first of all, does the field theory admit surface excitations with the anticipated fractional statistics; and second, are these excitations deconfined on the surface?

The question of statistics (in the Coulomb phase of the gauge theory, with $g^2$ small) for excitations near such a boundary was discussed for $k=1$ by Metlitski, Kane and Fisher.\cite{Metlitski13}  In Appendix \ref{SurfaceStatsApp} we discuss how to extend their result to the case of interest here, where $k>1$ and $g^2 \gg 1$.  Note that in a condensed phase, some of these statistical interactions (those involving electric and magnetic charges in multiples of those of the condensate) will be screened.  However, the statistical interactions of interest to us correspond to particles whose charges (and consequently whose statistical interactions) cannot be fully screened in the bulk.  
In this case (see Appendix \ref{SurfaceStatsApp}) 
we find that:
\begin{enumerate}
\item There is a surface contribution to the mutual statistics that exactly cancels the Berry phase interaction between the charge $(1,0)$ and the neutral monopole $(-k,1)$
\m  There is a surface contribution to the self-statistics of an electrically charged object, rendering the fundamental bosonic charge $(1,0)$ an anyon with exchange statistics of $- \pi /k$.  
\m  The neutral monopole remains a fermion (boson) near the surface for $k$ odd ($k$ even)
\end{enumerate}

 Hence pure bosons of the form $(n,0)$ acquire anyonic statistics when brought near the surface. One might worry that precisely these objects will be confined at the boundary between the $\theta =0$ and $\theta = 2 \pi k$ confined phases. However, the fact that the bulk Berry phase term between these objects and the neutral monopole is cancelled by a surface term suggests that though these objects are confined in the bulk,  very near to the surface they are deconfined.  

To understand this deconfinement, it is useful to consider what happens when we bring a charge near the surface.  Near a boundary across which $\theta$ changes discontinuously, a charged point particle will induce both a surface current (which can be described, in the Coulomb phase, by an image monopole on the other side of the boundary) and a surface charge (which can be described by an image charge)\cite{Qi09}.  
In the limit that $g^2 \rightarrow \infty$ the electric and magnetic image charges are given by
\be
q' = - q \ , \ \ \ m' = \frac{2 \pi}{ \theta }
\ee
In other words, the effect of bringing a charge near to the surface is to induce a net magnetic flux through the surface: at large $g^2$ the surface charge exactly cancels the bulk charge just below the surface, and the object has no net electric charge.  This is the intuitive reason that it can be deconfined.

One might worry that this charge neutralization also neutralizes the statistics.  However, there is an important distinction between the statistics of objects in which integer charges are bound to magnetic fluxes, and in which the charge associated with a magnetic flux is induced.  In the former case, for example, an object of charge $1$ and flux $\pi$ is a fermion; in the latter case it will be a semion.  This can be derived directly from the field theory; physically, it results from the relative angular momentum accumulated as the induced charge slowly builds up upon flux insertion\cite{Wilczek90}.  Hence though the object in question does not source a net electric charge, it does have fractional statistics. 

\subsection{Time reversal, obliquely confined phases, and the $\theta$ term}

We next turn to the question of time reversal.  In the field theory with bosonic sources $\theta$ is defined modulo $4 \pi$, such that for $\theta = 2 \pi k, k \in \mathbb{Z}$, the field theory is time-reversal invariant.  The fermionic toric code model discussed in \secref{FTCSec} is also manifestly time-reversal invariant. The Walker-Wang models corresponding to $k>1$, however, have Hamiltonians and ground-state wave-functions that are complex, and are not time-reversal symmetric.  We will briefly discuss why this does not present a contradiction.

To address this issue, 
let us review the conventional arguments for time reversal symmetry in the presence of a $\theta$ term.  
The basic idea is that if we consider a closed system (for example, the 3-torus $\{ x,y, z | 0 \leq x < L, 0 \leq y < L, 0 \leq z < L \}$ with periodic boundary conditions), then
\be
\frac{e^2}{ 32 \pi^2} \int \epsilon_{\mu \nu \rho \lambda} F^{\mu \nu} F^{\rho \lambda}  
\ee
 must be an integer.\cite{Callan76,Jackiw76,Wilczek87}  (One intuitive way to see this is that the electron wave-function must be single valued, which quantizes the magnetic flux through any spatial plane.\cite{Vazifeh10})  
Time reversal requires that a process that inserts a single flux quantum through the $xy$ plane, then changes $\oint A_z dz$ by $2 \pi$ (i.e. generates one quantum of electric field), should contribute the same overall phase to the action as its time-reversed conjugate, which is the same process with ${\bf B}$ reversed.  If $\theta$ is periodic modulo $2 \pi$, this criterion is satisfied provided that $\theta = \pi n$.  (In the presence of monopoles, one can further see that for bosons, time-reversal requires that $\theta = 2 \pi n$.\cite{Vishwanath13})

In these arguments, time-reversal symmetry is manifest when the partition function is computed on a closed manifold, and not as a local property of a Hamiltonian or ground-state wavefunction.  In this limited sense, our Walker-Wang models are also time-reversal symmetric: on closed spacetimes of the form $M^3 \times S^1$ the partition function is always $1$.\footnote{The Walker-Wang partition function can be complex; however, this requires a spacetime 4-manifold (such as $\mathbb{C P}^2$) with non-trivial signature, which we do not believe arises in physical situations}  

It is also worth emphasizing that on a manifold with boundary, the change in $\theta$ at the boundary necessarily breaks time-reversal (unless there is a gapless surface), and we would not expect time-reversal symmetry at the boundary for any value of $\theta$.  
Hence the fact that our lattice models admit chiral surface states is also not in conflict with our expectations from the field theory. 
Indeed the complex phases in the wave-function violate time-reversal in exactly the way that is required to allow the expected chirality to arise at the surface.  

Though it is certainly surprising that a model with a manifestly complex ground-state wave-function should describe a phase of the time-reversal invariant field theory with $\theta = 2 \pi k$, the Hamiltonian and ground states do satisfy the criterion of global time-reversal invariance required to be consistent with the field theory.

\section{Discussion and conclusions} \label{conclusion}

In this work, we have established that the obliquely confined phases of axion electrodynamics with integer $k=\theta/2\pi$ (and appropriate
dynamical matter fields) 
 can be described by a topological field theory (\ref{eq:bFbbJ2}) -- which can be expressed in Hamiltonian form by a Walker-Wang model.  
(We have discussed the specific example $k=1$ in the main text, but in Appendix \ref{s:quantisegenk} we show how these arguments apply in the general case).  We now discuss what conclusions about the phases realized by these lattice Hamiltonians can be drawn from this relation.

We begin by considering the phase diagram of axion electrodynamics for $k=1$.  
In the limit that $g^2 \rightarrow 0$ (when the U(1) gauge field is weakly fluctuating) and the dynamical  charge $2$ matter is heavy (so that it cannot condense and break this global U(1)), this field theory describes the electromagnetic response of the bosonic topological insulator.  As Ref. \onlinecite{Metlitski13} points out, one of the distinctive features of this phase is that the fundamental electrically neutral object (or `neutral monopole') of the compact U(1) gauge field is a fermion.  
 
Ref. \onlinecite{Metlitski13}  emphasizes that the fermionic nature of the neutral monopoles is necessary to cancel a statistical anomaly at the surface, where by Laughlin's argument we find that inserting a flux of $2 \pi$ creates a fermion.  Since these fermionic neutral monopoles must pair to condense, this  leads to $\mathbb{Z}_2^f$ topological order as we take the theory to strong coupling.  Hence the weak-coupling SPT becomes a topologically ordered phase at strong coupling.  

This self-same surface anomaly plays an important role in the Walker-Wang model: it means that we must include a trivial fermion, which will be deconfined in the bulk (leading to $\mathbb{Z}_2^f$ topological order).  In technical terms, the surface anomaly means that we must choose our anyon model not to be U(1)$_1$, but U(1)$_1 \times \mathbb{Z}_2^f$, which contains the vacuum and the ``trivial" fermion (which is the fermion created by inserting $2 \pi$ flux through the surface).  This result holds for general odd $k$: to build our Walker-Wang Hamiltonian, we must explicitly include the trivial fermion which, in the corresponding Laughlin state, arises from inserting flux $2 \pi k$ through the surface.  

Though we have focused exclusively on abelian anyon models, 
there are also non-abelian examples of anyon models with a similar surface statistical anomaly in which an appropriate flux insertion produces a ``trivial" fermion.  The corresponding 
Walker-Wang models also have bulk $\mathbb{Z}_2^f$ topological order, and can also presumably be viewed as the strong-coupling limit of weakly gauged symmetry-protected phases.  One example is the model based on SO(3)$_6$ used by Ref.~\onlinecite{Fidkowski13b} to discuss fermionic symmetry-protected phases.  (To do so they eliminate the $\mathbb{Z}_2^f$ topological order by introducing extra fermions into the problem).

Next, we consider the family of models with $k$ even.   It is instructive to scrutinize the surface states of these models: for even $k$ our surface states have the same anyon content as the bosonic Laughlin states at filling fraction $\nu=1/k$, which can be realized by a 2D U(1)$_k$ Chern-Simons gauge theory.  Hence for even $k$, in the absence of extra symmetries, our models are neither topologically ordered nor SPT\footnote{Recent work\cite{Kapustin13a,Kapustin13b,Gukov13} has shown that the phases considered in \secref{sec:confinedBTI} fall outside of the standard group cohomology classification, and such theories are `higher symmetry gauged' versions of `higher-symmetry' protected phases. Using results from the present work, the same two statements must be true of the $\mathbb{Z}_N$ Walker-Wang models.}. This suggests that the corresponding Walker-Wang Hamiltonians are adiabatically connected to the trivial confined phase with $\theta =0$.  

The field theory's phase diagram suggests a path whereby this can happen: it is well known that the trivial confined phase with $\theta =0$ is adiabatically connected to the trivial Higgs phase (in which a charge $1$ dynamical matter field has condensed)\cite{Fradkin79}.  In the Higgs phase, where magnetic monopoles are confined, $\theta$ can be changed arbitrarily without closing the gap.  After shifting $\theta$ by $4 \pi$, (as emphasized above) in the bulk of the system we simply obtain a new version of the trivial confined phase, in which our neutral monopole is described as the bound state of a monopole with two charges.  The thermodynamics, however, will be independent of this description; hence for $\theta = 4 \pi n, n\in \mathbb{Z}$, we should be able to adiabatically continue the system from the trivial Higgs phase to the appropriate obliquely confined phase.  Though (with appropriate boundary conditions) this will certainly close the gap at the surface, it should not require a bulk phase transition.

Notice that the above analysis is consistent with the observation\cite{Chen14} that it is possible to add a global symmetry to the Walker-Wang model with $k=2$ to obtain an SPT.   If the global symmetry is necessarily broken by condensing our (initially confined) charge $1$ matter field, then the path that we have identified to the trivial phase breaks the symmetry and the system may be an SPT.  (It is not clear, however, whether it is necessarily the case that the system {\it is} an SPT in this situation).  

It is worth emphasizing that, even for $k$ even, the Walker-Wang models have mathematically distinctive features.  Notably, their surfaces realize chiral topological gauge theories with discrete gauge group (and, in particular, zero correlation length).  In 2D, conversely, a complete classification of discrete topological gauge theories\cite{Dijkgraaf90} yields only achiral (or `doubled') theories like those seen in Levin-Wen type models\cite{Levin05,Lin14}. Indeed, it can be shown quite generally that the ground states of commuting projector 2D Hamiltonians are achiral in that they have vanishing central charge\cite{Lin14,Kitaev06}.


The same line of reasoning suggesting that all models with even $k$ are adiabatically connected to the trivial phase suggests that all models with odd $k$ are adiabatically connected to the model with $k=1$ (i.e. the fermionic Toric code).  This is consistent with an analysis of the robustness of the surface: because the 2D anyon model (including the fermion) can exist without the bulk, we can eliminate any excitations that are strictly confined to the surface by coupling a purely 2D system to the boundary of our 3D one.  The bulk fermion, on the other hand, obviously cannot be eliminated by such a surface term.
Somewhat perplexingly, in this case the phase diagram of the field theory offers no clear route adiabatically connecting these two phases.  Such a route must leave the system in a phase with $\mathbb{Z}_2$ topological order and a deconfined bulk {\it fermion}, which eliminates the possibility of passing through a Higgs phase (in which necessarily all deconfined charges are bosonic).  It also means that such a route cannot correspond to keeping $g^2$ large, due to the trivial obliquely confined phase at $k=2$.  We leave this as an interesting open question.

In summary, we have established that the obliquely confined phases of axion electrodynamics with integer $k=\theta/2\pi$ (and appropriately selected dynamical matter fields) are described by the Walker-Wang topological lattice models.  We have argued qualitatively that the spectra of the continuum models and the lattice models should be the same, in an appropriate long-distance limit; more importantly, we have supported these arguments by showing that the partition function of the gauge theory on a lattice can be viewed as the path integral of the Walker-Wang Hamiltonian.  As well as being pertinent to related work in the high-energy literature\cite{Kapustin13a,Kapustin13b,Gukov13}, this has allowed us to understand the relationship between the lattice models built from ``fermionic" quantum Hall states and SPT phases.  It also
suggests a route to adiabatically connect Hamiltonians constructed from bosonic quantum Hall states to the trivial confined phase.  

\acknowledgements

We would like to thank Steve Simon, Joseph Maciejko,  Olexei~I.~Motrunich, Ryan Thorngren and Max Metlitski for useful conversations.  FJB is supported by startup funds from the University of Minnesota. CVK acknowledges the financial support of the EPSRC and the Princeton Center for Theoretical Science.

\appendix

\section{Hamiltonian formulation of $bF+bb$ theory for $k\in\mathbb{Z},p=2k$}

\label{s:quantisegenk}
Here we identify a Hamiltonian for the $k\in\mathbb{Z},p=2k$ obliquely confined phases starting from the $bF+bb$ theory in \eqnref{eq:bFbbJ2}. First we express the partition function for the field
theory as a Trotter decomposition (\secref{ss:trotter}), and deduce
from this the effective Hamiltonian (\secref{ss:Heff}). The Hamiltonian obtained is that of the $\mathbb{Z}_{2k}^{(1)}$
Walker-Wang model\cite{vonkeyserlingk13a,Burnell13}. The special case $k=1,p=2$ corresponds to the Fermionic toric code
covered in \secref{s:quantiseoddk}. When $k>1$ is odd,
the surface has the topological order of a $\nu=1/k$ Laughlin state
and the bulk has Fermionic toric code $\mathbb{Z}^{f}_{2}$ topological
order. When $k$ is even, the surface has the topological
order of a $\nu=1/(k/2)$ Laughlin state coupled to a bulk Bosonic $\mathbb{Z}_2$ gauge theory.

\subsection{{Trotter decomposition}}

\label{ss:trotter}The $bF+bb$ Lagrangian density, in the presence of static charges and line defects ($j^0,\Sigma^{0i}$), can be
written 

\begin{equation}
-i b \wedge F-\frac{i2\pi}{p}b\wedge b- i j^{0} A_{0} -\Sigma^{0i} b_{0i} + \epsilon_V \mathcal{G}(e^{i\frac{2\pi}{p} j^{0}}) + \epsilon_{P}  \mathcal{F}(e^{i \Sigma^{0i}})  
\label{eq:bFbbgenk}
\end{equation}

where $F=dA$, and $A\in\left\{ 0,1,\ldots,p-1\right\} \frac{2\pi}{p}$,
and where $b_{0i},\frac{1}{2}\epsilon_{ijk}b_{jk}\in\left\{ 0,1,\ldots,p-1\right\} $. We remind the reader that $A$ lives on the edges of the 4D lattice, while $b$ lives on the faces of the dual lattice (\secref{ss:bFbbtoWW:Fieldtheory}). The fields $J^{0},\Sigma^{0i}\frac{p}{2\pi}$ take values in $\{0,1,2,\ldots,p-1\}$. The $\mathcal{F},\mathcal{G}$ functions endow these defects with an energy cost. As neither function has an explicit $A,b$ dependence, we will ignore them for now. To begin, sum out $A_0$ to get partition function

\begin{align}
\sum_{\{b\}}'\sum_{\{\tilde{A}\}}\exp\left[\sum_{r}(ib \wedge d\tilde{A}+\frac{i2\pi}{p}b\wedge b+\Sigma^{0i} b_{0i} )\right]\punc{,}
\end{align}
where $\tilde{A}_{\mu}$ contains only the remaining space-like components
of $A$, and where $\sum'$ is a sum over $b$ configurations constrained
by $\epsilon_{i j k } \partial_{i}b_{jk}=J_{0}\mod p$. Now
sum out the time-like components of $b$ (i.e. $b_{0k}$), to get
\begin{align}
 & \sum'_{\{A_{i},s_{i}\}}\exp\left[\sum_{r,i}-i\left(s_{r,i}-s_{r-\hat{0},i}\right)A_{r,i}\right]\nonumber \\
 & \times\prod_{(r,\hat{i},\hat{j})}\left[\sum_{l=0}^{p-1} e^{i l \Sigma^{0k}(r)} \left(\Phi_{(r,\hat{i},\hat{j})}\times T_{r+\hat{i}+\hat{j},k}T_{r-\hat{k}-\hat{0},k}\right)^{l}\right]\label{eq:genkpartition}
\end{align}
where we have relabeled the variables $b_{j k}(r)\rightarrow \epsilon_{i j k} s_{r,i}$
for the remaining space-like components of $b$, and $A_{i}(r)\rightarrow A_{r,i}$
for the remaining space-like components of $A$. The product in the
second line is over all space-like plaquettes at all times, $(r,\hat{i},\hat{j})$,
where $ij=12,23$ or $31$. Within this product, the index $k=1,2,3$
is fixed by the requirement $\epsilon_{ijk}=1$. We have used the
definitions

\begin{align*}
\Phi_{(r,\hat{i},\hat{j})}= & e^{iF_{ij}(r)}\\
T_{r,i}= & e^{\frac{i2\pi}{p}s_{r,i}}
\end{align*}
in \eqnref{eq:genkpartition}. In terms of the new $s_{i}$
variables, $\sum'$ means we only consider configurations of $s_{i}$
obeying vertex constraint $\partial_{i}s_{r,i}=J_{0}(r)\mod p$ at
each vertex $r$ of the lattice within each time slice. To
form the effective Hamiltonian, we express the partition function
\eqnref{eq:genkpartition} as a product over space-like slices indexed
by time co-ordinate $r^{0}=n$

\begin{align}
&\sum'_{\{A_{e},s_{e}\}}\prod_{n}\left\{ \exp\left[\sum_{e}-i\left(s_{e}^{(n)}-s_{e}^{(n-1)}\right)A_{e}^{(n)})\right]\right.\nonumber \\
&\left.\times \prod_{P}\left[\sum_{l=0}^{p-1} e^{i l \Sigma_{P}}\left(\Phi_{P}^{(n)}\times T_{O(P)}^{(n)}T_{U(P)}^{(n-1)}\right)^{l}\right]\right\}
 \label{eq:pretrotter}
\end{align}

where the $P$ labels oriented $12,23,31$ plaquettes on a 3D cubic
lattice. On the other hand, $e,V$ label edges and vertices
respectively of the cubic lattice. The edges are by default oriented
in the $+\hat{1},+\hat{2},+\hat{3}$ directions. In Eq.~\ref{eq:pretrotter}, $\Phi_{P}^{(n)}, \Sigma_{P}$ is shorthand for the value of $\Phi,\Sigma$ on a plaquette
$P$ of the cubic lattice at time slice $n$; note $\Sigma_{P}$ does not require a time slice co-ordinate as it represents a static line defect. $T_{O(P)}^{(n)},T_{U(P)}^{(n-1)}$ denote the values of $T$ on two
edges $O(P),U(P)$ at at time-slices $n,n-1$ respectively. If $P$
is the $ij\in{12,23,31}$ plaquette on the 3D cubic lattice at vertex
$\vec{r}$, then $O(P)$ is the edge $(\vec{r}+\hat{i}+\hat{j},\hat{k})$
while $U(P)$ is the edge $(\vec{r}-\hat{k},\hat{k})$, where $k$
is uniquely specified by $\epsilon_{hij}=1$. $s_{e}^{(n)},A_{e}^{(n)}$
denote the values of $s,A$ on the space-like edge $e$, at time-slice
$n$. 

Now notice that the second line of \eqnref{eq:pretrotter} imposes a constraint

\be
\mathcal{B}^{(n)}_P\equiv \Phi_{P}^{(n)}\times T_{O(P)}^{(n)}T_{U(P)}^{(n-1)} =e^{-i\Sigma_{P}}
\ee
We remind the reader that the $\Phi,T$ depend only on $A,s$ respectively. Moreover, the Gauss constraint which holds at each vertex $V$ and time slice $n$, can be written

\begin{equation}\label{eq:gaussHam}
\mathcal{Q}^{(n)}_V \equiv \prod_{e\in s(V)}\db{T^{(n)} _{e}}^{\nu_{e,V}}=e^{\frac{i2\pi}{p}J_{0}(V)}\punc{,}
\end{equation}

where $s(V)$ denotes the six edges attached to a vertex $V$, and $\nu_{e,V}=\pm1$ depending on whether $e$ is oriented towards or away from $V$. Using the constraints to remove all explicit dependence on the static sources $J^0,\Sigma^{0i}$ and, re-introducing the $\mathcal{F},\mathcal{G}$ functions from \eqnref{eq:bFbbgenk} gives partition function

\begin{align}
&\sum'_{\{A_{e},s_{e}\}}\prod_{n}\left\{ \exp\left[\sum_{e}-i\left(s_{e}^{(n)}-s_{e}^{(n-1)}\right)A_{e}^{(n)})\right]\right.\nonumber \\
&\left.\times \exp\ds{\epsilon_v \sum_V\mathcal{G}\db{\mathcal{Q}^{(n)}_V} + \epsilon_p \sum_P \mathcal{F}\db{\mathcal{B}^{(n)}_P } }\right\}
 \label{eq:lastpretrotter}
\end{align}

\subsection{{The effective Hamiltonian}\label{ss:Heff} }

Having expressed the classical partition function as \eqnref{eq:pretrotter},
it is relatively straightforward to rewrite it as $\tr e^{-\hat{H}_{\text{eff}}\tau}$
where $\hat{H}_{\text{eff}}$ is a 3D quantum Hamiltonian. In \eqnref{eq:pretrotter}, $s_{i},\frac{pA_{i}}{2\pi}=0,1,2,\dots p-1$
are classical variables, but they will become quantum mechanical operators
$\hat{s}_{i},\hat{A}_{i}$ when we move to the Hamiltonian formulation
of this problem. They obey commutation relations

\begin{equation}
\left[\hat{A}_{r',j'},\hat{s}_{r,j}\right]=i\delta_{r,r'}\delta_{j,j'}\mod p\punc{,}\label{eq:comm}
\end{equation}

and the local Hilbert space on an edge $e$ of the cubic lattice has
two obvious bases with $p$ elements

\begin{align*}
\hat{A}_{e}\left|A_{e}\right\rangle =A_{e}\left|A_{e}\right\rangle , & A_{e}\in\frac{\pi}{k}\left\{ 0,1,\ldots,p-1\right\} \\
\hat{s}_{e}\left|s_{e}\right\rangle =s_{e}\left|s_{e}\right\rangle , & s_{e}\in\left\{ 0,1,\ldots, p-1\right\} \punc{.}
\end{align*}

It is convenient to construct operators $\hat{W}_{e}=e^{i\hat{A}_{e}},\hat{T}_{e}=e^{\frac{i 2\pi}{p}\hat{s}_{e}}$,
and pick a phase convention for the bases such that

\begin{align*}
\hat{W}_{e}\left|s_{e}\right\rangle = & \left|s_{e}+1\right\rangle \\
\hat{T}_{e}\left|A_{e}\right\rangle = & \left|A_{e}-\frac{2\pi}{p}\right\rangle \punc{.}
\end{align*}

We are now ready to introduce these quantum mechanical operators into
\eqnref{eq:pretrotter}. It follows from Eq.~\ref{eq:comm} that
\[
\left\langle s_{e}'\right|\left.A_{e}\right\rangle \left\langle A_{e}\right|\left.s_{e}\right\rangle =e^{iA_{e}\left(s_{e}-s_{e}'\right)}
\]

which allows us to re-express the partition function as

\begin{align*}
 & \sum'_{\{A_{e},s_{e}\}}\prod_{n=r^{0}}\left\{ \left[\prod_{e}\left\langle s_{e}^{(n)}\right|\left.A_{e}\right\rangle \left\langle A_{e}\right|\left.s_{e}^{(n-1)}\right\rangle \right]\right.\\
 & \left.  \times \exp\ds{\epsilon_v \sum_V\mathcal{G}\db{\mathcal{Q}^{(n)}_V} + \epsilon_p \sum_P \mathcal{F}\db{\mathcal{B}^{(n)}_P } }\right\} 
\end{align*}

We can collect the products over bras and kets to form

\begin{align*}
 & \sum'_{\{A_{e},s_{e}\}}\!\prod_{n}\!\!\left\{ \!\left\langle \{s^{(n)}\}\right|\prod_{P} e^{\epsilon_p \sum_P \mathcal{F}\db{\Phi^{(n)}_{P} \times T_{O(P)}^{(n)}T_{U(P)}^{(n-1)} } }\right.\\
 & \left.\times\left|\{A^{(n)}\}\right\rangle \left\langle \{A^{(n)}\}\right|\left.\{s^{(n-1)}\}\right\rangle \times e^{\epsilon_v \sum_V \mathcal{G}\db{ \mathcal{Q}^{(n)}_V }} \right\} 
\end{align*}

where $\left|\{s^{(n)}\}\right\rangle =\otimes_{e}\left|s_{e}^{(n)}\right\rangle $
and $\left|\{A^{(n)}\}\right\rangle =\otimes_{e}\left|A_{e}^{(n)}\right\rangle $.
We have also moved the $\mathcal{F}$ terms between the kets. We can now replace $\Phi^{(n)}_{P}$ by the operator $\hat{\Phi}_{P}$ by substituting $A_{e}^{(n)}\rightarrow\hat{A}_{e}$

\begin{align*}
 & \sum'_{\{A_{e},s_{e}\}}\!\prod_{n}\!\!\left\{ \!\left\langle \{s^{(n)}\}\right|\prod_{P} e^{\epsilon_p \sum_P \mathcal{F}\db{\hat{\Phi}_{P}\times T_{O(P)}^{(n)}T_{U(P)}^{(n-1)} } }\right.\\
 & \left.\times\left|\{A^{(n)}\}\right\rangle \left\langle \{A^{(n)}\}\right|\left.\{s^{(n-1)}\}\right\rangle \times e^{\epsilon_v \sum_V \mathcal{G}\db{ \mathcal{Q}^{(n)}_V }} \right\} \punc{.}
\end{align*}

 Summing over
$A_{e}$, and using $\hat{I}=\sum_{A_{e}^{(n)}}\left|\{A^{(n)}\}\right\rangle \left\langle \{A^{(n)}\}\right|$
gives

\begin{align}
\sum'_{\{s_{e}\}} \!&\prod_{n}\!  \!\left\langle \{s^{(n)}\}\right|\prod_{P} e^{\epsilon_p \sum_P \mathcal{F}\db{\hat{\Phi}_{P}\times T_{O(P)}^{(n)}T_{U(P)}^{(n-1)} } }  \left|\{s^{(n-1)}\}\!\right\rangle \nonumber\n
&\times e^{\epsilon_v \sum_V \mathcal{G}\db{ \mathcal{Q}^{(n)}_V }} \label{eq:prefinalpartition}
\end{align}
where the $T$ variables are currently still numbers depending on the $s$ variables, rather than
operators. The product between the kets in \eqnref{eq:prefinalpartition} Taylor
expands to a sum of expressions of the form

\[
\left\langle \{s^{(n)}\}\right|\prod_{P\in S}\left[T_{O(P)}^{(n)}T_{U(P)}^{(n-1)}\!\hat{\Phi}_{P}\right]^{j_{P}}\left|\{s^{(n-1)}\}\right\rangle 
\]

where $P=(\vec{r},\hat{i},\hat{j})$. Here the $j_{P}'s$
are some integers and $S$ is some set of space-like plaquettes. Now,
this can be rewritten as

\begin{equation}
\left\langle \{s^{(n)}\}\right|\!\!\left(\prod_{P\in S}\hat{T}_{O(P)}^{j_{P}}\right)\!\!\left(\prod_{P\in S}\hat{\Phi}_{P}^{j_{P}}\right)\!\!\left(\prod_{P\in S}\hat{T}_{U(P)}^{j_{P}}\right)\!\!\left|\{s^{(n-1)}\}\right\rangle \label{eq:timeordering0}
\end{equation}

where we have replaced the $s,T=e^{\frac{i2\pi}{p}s}$ variables by
$\hat{s},\hat{T}=e^{\frac{i2\pi}{p}\hat{s}}$ operators respectively.
This can be arranged to

\begin{equation}
\left\langle \{s^{(n)}\}\right|\prod_{P\in S}\left[\hat{T}_{O(P)}\hat{T}_{U(P)}\hat{\Phi}_{P}\right]^{j_{P}}\left|\{s^{(n-1)}\}\right\rangle \label{eq:timeordering}
\end{equation}

One might worry about the phases accrued in permuting the $\hat{T},\hat{\Phi}$
operators to go between \eqnref{eq:timeordering0} and \eqnref{eq:timeordering}.
If $e\in\partial P$ then note that

\[
\hat{T}_{e}\hat{\Phi}_{P}=\hat{\Phi}_{P}\hat{T}_{e}e^{\frac{i2\pi}{p}\nu_{e,P}}
\]

where $\nu_{e,P}$ is the orientation of $e$ in the boundary of $P$.
But all of these phases cancel. Why is this? To rearrange Eq.~\ref{eq:timeordering}
to form Eq.~\ref{eq:timeordering0} we need to move every $\hat{T}_{O}$
operator to the left of the product and every $\hat{T}_{U}$ operator
to the right of the product. Consider two plaquettes $P,Q$. If $e$
is in the boundary of $P$ (i.e. $e\in\partial P$) with orientation
$\nu$ but is also an O edge of $Q$ ($e=O(Q)$, it follows by inspection
of \figref{fig:plaquette} that there exists a unique $f\in\partial Q$
with $\nu_{f,Q}=\nu$ such that $f$ is a U edge of $p$ i.e., $f=U(P)$.
This tells us the following. We need to move $\hat{T}_{e}$ to the
left of Eq.~\ref{eq:timeordering} because $e$ is an O edge, and
we also need to move $\hat{T}_{f}$ to the right because it is a U
edge. If, in Eq.~\ref{eq:timeordering}, $Q$ lies to the left of
$P$, $\hat{T}_{e}$ will not meet the plaquette operator at $Q$
and $\hat{T}_{f}$ will not meet the plaquette operator at $P$. If
$Q$ lies to the right of $P$, $\hat{T}_{e}$ encounters the $P$
plaquette operator and it will accrue a net $e^{-\frac{i2\pi}{p}\nu_{e,P}j_{P}j_{Q}}$
sign because $e\in\partial P$. However as $Q$ is to the right of
$P$, we will also need to move the U edge $f$ of plaquette $P$
to the right of Eq.~\ref{eq:timeordering}. It will need to move
through the $Q$ plaquette operator, and because $f\in\partial P$
it will acquire a $e^{\frac{i2\pi}{p}\nu_{f,Q}j_{P}j_{Q}}$ phase as
it does so. In either case, there is no net phase because $\nu_{e,P}=\nu_{f,Q}$.
We can proceed inductively, to prove that Eq.~\ref{eq:timeordering}
equals Eq.~\ref{eq:timeordering0}. Having shown this, we can go
back to our expression for the partition function and recast it as

\begin{align}
\sum'_{\{s_{e}\}} \!&\prod_{n}\!  \!\left\langle \{s^{(n)}\}\right|\prod_{P} e^{\epsilon_p \sum_P \mathcal{F}\db{\hat{\Phi}_{P}\times\hat{T}_{U(P)}\hat{T}_{O(P)}} }  \left|\{s^{(n-1)}\}\!\right\rangle \nonumber\n
&\times e^{\epsilon_v \sum_V \mathcal{G}\db{ \mathcal{Q}^{(n)}_V }} \label{eq:prefinalpartition-1}
\end{align}

We can now move the  $ \mathcal{Q}^{(n)}_V$ fields in-between the kets and replace them by operators

\begin{equation}\label{eq:gaussHam}
\hat{Q}'_V \equiv \prod_{e\in s(V)}\hat{T}_{e,\nu_{e,V}}=e^{\frac{i2\pi}{p}J_{0}(V)}\punc{.}
\end{equation}

where $s(V)$ are the six edges incoming to vertex $V$, and $\nu_{e,V}=\pm1$ depending on whether $e$ is oriented towards/away from $V$ respectively. This allow us to rewrite the partition function as
\begin{align*}
\sum'_{\{s_{e}\}}\prod_{n}\left\{ \left\langle \{s^{(n)}\}\right|e^{-\hat{H}_{\text{{eff}}}\Delta\tau}\left|\{s^{(n-1)}\}\right\rangle \right\} 
\end{align*}

where we define $\hat{H}_{\text{{eff}}}$ implicitly though

\begin{equation}\label{eq:sepHeff}
  e^{-\hat{H}_{\text{{eff}}}\Delta\tau}=    \exp \ds{ \epsilon_P \sum_{P}   \mathcal{F}\db{\hat{B}'_{P}} }  \exp \ds{\epsilon_V \sum_{V} \mathcal{G}\db{ \hat{Q}'_V}}
\end{equation}
and we defined  
\begin{align*}
&\hat{B}'_{P}\equiv \left(\hat{\Phi}_{P}\hat{\Theta}_{P}\right)^{l}\\
&\hat{\Phi}_{P}\equiv  \prod_{e=(\vec{r},\hat{i})\in\partial P}\hat{W}_{e}^{\nu_{e,P}}\\
&\hat{\Theta}_{P}\equiv  \hat{T}_{O(P)}\hat{T}_{U(P)} \punc{.}
\end{align*}

Here $O(P),U(P)$ are shorthand for the edges $(r+\hat{i}+\hat{j},\hat{k}),(r-\hat{k},\hat{k})$
transverse to the plaquette $P=(x,\hat{i},\hat{j})$ where $\epsilon_{hij}=1$. Also, $\nu_{e,P}$ is the orientation of the edge $e=(\vec{r},\hat{i})$ in the plaquette boundary $\partial P$.  It is straightforward to verify that the operators $\hat{Q}'_{V}$ and $\hat{B}'_{P}$ commute amongst themselves so we can combine the exponents in \eqnref{eq:sepHeff} to get

\begin{equation}\label{eq:WWprefinal}
\hat{H}_{\text{{eff}}}\Delta\tau= -\epsilon_{V} \sum_{V}  \mathcal{G}\db{ \hat{Q}'_V} - \epsilon_{P} \sum_{P} \mathcal{F}\db{\hat{B}'_{P} }  
\end{equation}
We now specify functions $\mathcal{G},\mathcal{F}$. We want both functions to penalize configurations with non-vanishing $\Sigma^{0i}$ and $J^{0}$. A simple option is to take $\mathcal{G}=\mathcal{F}( \kappa )= p\delta_{\kappa,1}-1$, which penalizes all non-vanishing defects equally, and leads to the Walker-Wang Hamiltonian for the category $\mathbb{Z}_{p}^{(1)}$\cite{Walker12,vonkeyserlingk13a}

\begin{equation}\label{eq:WWfinal}
H\Delta\tau=-\epsilon_v \sum_V \hat{Q}_{V}-\epsilon_p \sum_{P}\hat{B}_{P}\punc{.}
\end{equation}

where 
\begin{align}\label{eq:stabs}
\hat{ Q }_{V}&=\sum_{l=1}^{p-1}\db{\hat{ Q }'_{V}}^{l}\n
\hat{B}_{P}&=\sum_{l=1}^{p-1} \db{\hat{B}'_{P}}^{l} \punc{.}
\end{align}

The ground states and topological properties of excitations do not depend on the precise form of 
 $\mathcal{F}, \mathcal{G}$, provided these functions are maximized for configurations with $\hat{Q}'_V=\hat{B}'_P=1$.

\section{$k=\text{even}, p=k$}
	\label{sec:kevenpk}

In the previous section, we showed that the $k=\text{even}, p=2k$ $bF+bb$ theory is precisely the $\mathbb{Z}_{2k}^{(1)}$ Walker-Wang model.  
However, our heuristic discussion in \secref{sec:confinedBTI} also suggests that the $\mathbb{Z}_k^{(1/2)}$ Walker-Wang model (constructed from the $U(1)_k$ Chern-Simons theory) is described by the same field theory with $k=\text{even}, p=k$. 
In this case, however, we run into a technical obstacle in making this correspondence rigorous.
In particular, for a lattice version of the field theory to be equivalent to the $\mathbb{Z}_k^{(1/2)}$ Walker-Wang model,  we would expect the action to be periodic modulo $2\pi$ under $b\rightarrow b+k\eta$ for arbitrary integer two-form $\eta$, reflecting the fact that label $k$ strings in the WW ground state are trivial. However, this fails to be true if we use the definition of $b\wedge b$ in \eqnref{eq:bblattice}, meaning that the proof given above cannot be applied directly in this case. 

Ref.~\onlinecite{Kapustin13a} resolve this issue on a simplicial manifold by using the `Pontryagin square' operation to define  $b\wedge b$.   The Pontryagin square is a simple extension of the standard cup product $b \cup b \rightarrow b\cup b + b\cup_1 db$ in simplicial cohomology.  (The cup product is the simplicial manifold analogue of the wedge product that we use here).
We expect that with a similar refinement of the wedge product on the hypercubic lattice \eqnref{eq:bblattice}, one could use the methods of \appref{s:quantisegenk} to derive the $\mathbb{Z}_k^{(1/2)}$ Walker-Wang models from the $p=k$ even $bF+bb$ action.

\section{Relation to previous work} \label{SVApp}

It is worth noting that the BTI -- and the corresponding strong-coupled FTC phase -- can be obtained from a Walker-Wang model in a different way.  As observed in Ref.~\onlinecite{Vishwanath13}, the BTI is described by the (fully gapped) field theory
\be \label{VSBF}
\mac{L} \!= \!\frac{1}{2 \pi} \epsilon^{\mu\nu\rho\lambda} \! \left( \! B^{(1)}_{\mu \nu} \partial_\rho a^{(1)}_\lambda\! +  \!B^{(2)}_{\mu \nu} \partial_\rho a^{(2)}_\lambda \right) + \frac{\theta \epsilon^{\mu\nu\rho\lambda}}{4 \pi^2} \partial_\mu a^{(1)}_\nu  \partial_\rho a^{(2)}_\lambda 
\ee
in which the sources of $a^{(1)}$ and $a^{(2)}$ have the same charge under a global U(1) symmetry (or equivalently, under a non-dynamical electromagnetic gauge field).  
The field theory action (\ref{VSBF}) is also invariant under the time-reversal transformation
\ba
B^{(I)}_{0i} \rightarrow - B^{(I)}_{0i} \ , \ \ \ \ \ \ B^{(I)}_{ij} \rightarrow  B^{(I)}_{ij} \n 
a^{(I)}_0 \rightarrow a^{(I)}_0 \ , \ \ \ \ \ \ a^{(I)}_i \rightarrow -a^{(I)}_i
\ea
It represents a phase in which there are no deconfined point particles in the bulk, but whose surface is either gapless, symmetry-breaking, or contains $\mathbb{Z}_2$ topological order.  This is because on the surface, the two types of vortices both carry a U(1) charge of 1/2, and transform into one another under time reversal -- hence a condensate of vortices of one type results in a state that breaks $T$ and has a surface Hall conductivity $\sigma_{xy} = \pm 1$.   Since vortices of species $1$ are charged under species $2$ and vice versa, a bound state of these two vortices, while time-reversal invariant, is a fermion, and hence these objects must pair in order to condense, resulting in the $\zt$ surface topological order.  

As Ref.~\onlinecite{Vishwanath13} suggested, the field theory (\ref{VSBF}) can be realized as a Walker-Wang model, whose detailed description we will provide presently.  One might therefore conclude that this is the Walker-Wang model that one should consider in the context of the BTI.  However, in order to do so, one would have to incorporate the global U(1) symmetry into the lattice model.  Again, we will discuss in more detail below how this can be done; however, the important point is that the Walker-Wang model per se does not have the global U(1) symmetry and hence is not a BTI.  (Nor does it represent another member of the cohomology classification of Ref. \onlinecite{Chen13}, since the $\zt$ gauge theory at its surface is a perfectly legitimate 2D topological order).

\subsection{A Walker-Wang model for Eq. \ref{VSBF} }

The Walker-Wang model that captures the physics of the field theory (\ref{VSBF}) has the Hamiltonian
\be \label{Ham}
H = - \sum_V A_V - \sum_P B_P
\ee
with
\be \label{Av}
A_V=  \prod_{i \in *V}\sigma^z_i + \prod_{i \in *V} \tau^z_i
\ee
and
\be
B_P = B_P^{(e)} + B_P^{(m)} + B_P^{(e)} B_P^{(m)}
\ee
with 
\begin{eqnarray} \label{eq:bpe2}
B_P^{(e)}= \tau_{i_U}^z \prod_{ i \in \partial P } \sigma^x_i \ , \ \ \ 
B_P^{(m)}=\sigma_{i_O}^z  \prod_{ i \in \partial P } \tau^x_i.
\end{eqnarray}
where the $O$ and $U$ edges are shown in Fig. \ref{fig:plaquette}.
By the general results of Ref.~\cite{WWUs}, this Hamiltonian has no deconfined excitations in the bulk.  Further, it  is time-reversal invariant (the Hamiltonian being explicitly real) with a $\zt$ surface topological order, matching that expected for the bosonic topological insulator.\footnote{ Time reversal acts trivially on $e$ and $m$ here, since the condensate is of paired vortices, so the fact that one transformed into the other under $T$ before condensation is no longer material.}  

To turn this into a BTI, however, we must introduce a global U(1) symmetry under which $e$ and $m$ have charge $1/2$ (mod 1).  This symmetry plays an essential role in the symmetry protection of the surface state: as is well known, in the $\mathbb{Z}_2$ topologically ordered surface theory we can condense $e$ or $m$ to obtain a  trivial insulator at the surface.  However, in the BTI $e$ and $m$ are charged under U(1), so this trivial surface state breaks the U(1) symmetry.   Their composite $em$ can be U(1) neutral, but as it is a fermion, it cannot condense.  Hence only pairs of the $em$ excitations can condense without breaking the U(1) symmetry -- which obviously does not change the surface topological order.  

\subsubsection{Incorporating the U(1) symmetry}

We next turn to the question of how to incorporate the U(1) symmetry into our model, such that the $e$ and $m$ vertex defects have charge $\pm 1/2$, and their composite $em$ is neutral.  Here we will take the view that ``fundamental'' objects carry integer U(1) charge, and that any half-integer charges must arise due to collective effects.  We will therefore use a construction very similar to that introduced by Ref.~\onlinecite{Levin11}, which creates a model with vertex excitations carrying fractional (conserved) U(1) charge.  

In addition to the $\zt$ degrees of freedom that are required to describe the BF theory, we will include boson creation operators $e^{ i \theta_V}$ at each vertex $V$ on the lattice.  We will work in the number-phase representation, where the number of bosons at each site can be positive or negative (which we may interpret as meaning that there is some fixed but large density $\on$ at each site, and we measure the number $n_V$ of bosons on each site relative to this mean).   On every edge of the lattice, we will also include a single orbital on which these bosons can sit, with an extremely large Mott repulsion term, such that the state of each link can be described by the eigenvalues of $\sigma^z, \tau^z$ and $\alpha^z$, where 
\be
\alpha^z_{VV'} = 1-2 n_{VV'} 
\ee
 is a new 2-state variable indicating whether the orbital's occupancy $n_{VV'}= 0$ or $1$ (all other states being excluded from our Hilbert space as they are too high in energy).  We will denote by $\alpha^{+}_{VV'}$ ($\alpha^{-}_{VV'}$) the operator that raises (lowers) the number of bosons on the edge $VV'$ by $1$ {\it within the low-energy Hilbert space} where $n_{VV'} = 0,1$.   We therefore have 
 \be
 \alpha^+ _{VV'} = \begin{pmatrix} 0 & 1 \\
 0 & 0 \\
 \end{pmatrix}
  \ , \ \ \ \ \ 
   \alpha^- _{VV'} = \begin{pmatrix} 0 & 0 \\
 1 & 0 \\
  \end{pmatrix}
 \ee
  and $\alpha^x_{VV'} \equiv \alpha^+{VV'} + \alpha^-_{VV'}$ flips the sign of the eigenvalue of $\alpha^z_{VV'}$.  

Since each edge is shared by 2 sites, the operator that determines the charge localized near a given site is:
\be
q_V = n_V + \frac{1}{2} \sum_{<VV'>} n_{VV'}
\ee
It follows that the operator 
\be
\alpha^+_{VV'}e^{ -i  \theta_{V'}}
\ee
creates a dipolar charge distribution,
\be
\langle q_V \rangle = - \langle q_{V'} \rangle = \frac{1}{2}
\ee
relative to the equilibrium.  

Clearly, one way to obtain the correct charges for the $e$ and $m$ excitations (which are confined in the bulk, and deconfined at the surface) is to force edges on which  $\tau^z_{VV'}$ or $\sigma^z_{VV'}=-1$ to have a dipole moment such that $\langle q_V \rangle$ and $\langle q_{V'} \rangle = \pm 1/2$.   
In other words, we should impose the constraint
\be \label{Vconstr}
 n_V + \frac{1}{2} \sum_{<VV'>} n_{VV'} 
= \frac{1}{4} \left( 1 - \prod_{V'} \sigma^z_{VV'} \tau^z_{VV'}\right )
\ee
where, as usual, the product and sum run over vertices $V'$ that are neighbors of $V$.  

To make our plaquette operator compatible with this constraint, we will replace $\sigma^x_{VV'}, \tau^x_{VV'}$ in our Hamiltonian by the modified operators
\ba
\ts^x_{VV'} &=&\sigma^x_{VV'}  \alpha^+_{VV'}  e^{ -i \theta_{V}} + \sigma^x_{VV'}  \alpha^-_{VV'}  e^{ i \theta_{V'}}
\n
\ta^x_{VV'} & =& \tau^x_{VV'}  \alpha^+_{VV'}  e^{ -i \theta_{V'}} +\tau^x_{VV'}  \alpha^-_{VV'}  e^{ i \theta_{V}}
\ea
We note that
\be \label{ChargeCom}
\left[ \ts^x_{VV'}, q_{V''} \right ] = \frac{1}{2}\ts^x_{VV'} \left(  \delta_{V'', V'}  - \delta_{V'', V} \right)
\ee
Our new plaquette operators are
\begin{eqnarray} \label{eq:bpe}
B_P^{(e)}= \frac{1}{2} \tau_{i_U}^z \left ( \prod_{ i \in \partial P }   \ts^x_i + \prod_{ i \in \partial P } (  \ts^x_i)^\dag \right ) \\
B_P^{(m)}=\frac{1}{2} \sigma_{i_O}^z \left (  \prod_{ i \in \partial P } \ta^x_i + \prod_{ i \in \partial P } (  \ta^x_i)^\dag \right ) .
\end{eqnarray}
Their product is unchanged from the original model, since
\be
\ts^x_{VV'} \ta^x_{VV'}  = \sigma^x_{VV'}\tau^x_{VV'} \left \{ \alpha^+_{VV'} ,  \alpha^-_{VV'} 
 \right \} =  \sigma^x_{VV'}\tau^x_{VV'} 
\ee
Eq. \ref{ChargeCom} ensures that any operator in which $\ts^x$ or $\ta^x$ act on a closed loop of edges -- and in particular, the operators $B_P^{(e)}$ and $B_P^{(m)}$ --  commute with the left-hand side of Eq. (\ref{Vconstr}).  
The commutator with the right-hand side is also trivial since each plaquette term acts with $\sigma^x$ and $\tau^x$ on an even number of edges at each vertex.  (Indeed, $B_P^{(e)}$ and $B_P^{(m)}$ commute with $\prod_{V'} \sigma^z_{VV'}$ and $\prod_{V'} \tau^z_{VV'}$ individually).  Hence our modified plaquette operator commutes with both sides of the constraint (\ref{Vconstr}), and we may consistently work in a restricted Hilbert space where this constraint is obeyed.  

Let us now consider the nature of the ground states of our model so far.  In the ground sates, $ \prod_{V'} \sigma^z_{VV'} \tau^z_{VV'} =1$ so that $ - n_V = \frac{1}{2} \sum_{VV'} n_{VV'}$.  There will be no net charge about any vertex, and the number of edges entering each vertex on which $n_{VV'} =1$ must be even.  Further, the plaquette terms ensure that the ground state is a ``loop soup" over all possible trivalent graphs involving $e$, $m$, and $em$.  Such a state can be built, for example, by taking
\be
|\Psi_0 \rangle = e^{ \sum_P B_P }  |\sigma^z_{VV'} = 1, \tau^z_{VV'} = 1, n_V =0, n_{VV' } =0 \rangle
\ee
However, this is not the only possible choice -- clearly acting with $e^{\sum_P B_P}$ on {\it any} state of the form
\be
|\Psi_{ \{ n_V \}  }  \rangle = e^{ \sum_P B_P }  |\sigma^z_{VV'} = 1, \tau^z_{VV'} = 1,  - n_V = \frac{1}{2} \sum_{VV'} n_{VV'} \rangle
\ee
is actually a ground state.  To lift this degeneracy and ensure that our model for the BTI has a unique ground state (on a closed 3D system), we add a term of the form
\be \label{dh2}
\delta H = - \sum_{VV'}  \sigma^z_{VV'} \tau^z_{VV'} \left(\frac{1}{2}- n_{VV'} \right )
\ee
to our Hamiltonian.  This favors configurations where $n_{VV'} = 1$ when $ \sigma^z_{VV'} \tau^z_{VV'} = -1$, and $n_{VV'} =0$ if $\sigma^z_{VV'} \tau^z_{VV'} =1$.

Next, consider gauging the U(1) sector of this theory.  The gauge invariant boson kinetic terms are then
\ba
b^\dag_{V} e^{i A_{1,VV'} } b_{VV'}, \ \ \ b_{V} e^{- i A_{1,VV'} } b^\dag_{VV'}  \n
b^\dag_{V'} e^{- i A_{2,VV'} } b_{VV'}, \ \ \ b_{V'} e^{ i A_{2,VV'} } b^\dag_{VV'} 
\ea
Because charged bosons can live both on the vertices and on the edges, we have introduced two gauge fields ($A_{1,VV'}$ and $A_{2, VV'}$) on each edge. 

Generically, we should introduce $E^2$ and $B^2$ terms for our U(1) gauge theory on the lattice.  However, since we are principally interested in the confining limit (where the coefficient of the $E^2$ term is large, while that of the $B^2$ term is negligibly small), let us focus on the possible electric field configurations in our model.  
Since $n_{VV'} = 0,1$, there are two possibilities for the electric flux on each edge: either $E_{1,VV'} = E_{2, VV'}$ if $n_{VV'} =0$, or $E_{1,VV'} = E_{2, VV'} + 1$ if $n_{VV'} =1$.  At each vertex, the constraint
\be
\sum_{V'} E_{VV'} = n_V
\ee
must also be satisfied.  Notice also that when our Hamiltonian acts to change  $ \sigma^z_{VV'} \tau^z_{VV'}$ on edge $VV'$, this either adds or removes a boson from the center of the link, and hence changes $|E_{1,VV'} +E_{2,VV'}|$ by $1$. 


Let us consider what happens if we add to our model a term that penalizes non-vanishing electric flux.  (Such a term is certainly present in the confining phase).  The simplest way to do this is to consider adding a small $ \delta E^2$ term (and no $B^2$ term) to the U(1) lattice Hamiltonian.  The ground states of our model will then be those that minimize $E^2$ on each edge, all other things being equal.  Thus even at small $\delta$, we expect that with such a term we will find $E=0$ except on edges where $\prod \sigma^z \tau^z =-1$, where 
\be
E_1^2 + E_2^2 \geq 1
\ee

This suggests that we can understand the phase where U(1) monopoles have proliferated by considering
\be
H = - \sum_V A_V - \sum_P B_P - h \sum_e \tau^z_e \sigma^z_e
\ee
A general framework for understanding the ``condensed'' limit (when $h$ becomes large) of such models was laid out on Ref. \onlinecite{Burnell13}.  In the present case, the term proportional to $h$ penalizes edges for which $\tau^z$ {\it or} $\sigma^z =-1$, but not edges on which $\tau^z=\sigma^z =-1$.  We can therefore obtain a solvable model describing the large $h$ limit by restricting our attention to edges for which $\tau^z = \sigma^z$, and keeping only the product $B_P^{(e)} B_P^{(m)}$ (which does not take states out of this subspace) in the plaquette term.  The resulting Hamiltonian is precisely that of the fermionic Toric code.   

Thus if we incorporate the U(1) symmetry in the Walker-Wang model and gauge it in such a way that lines of U(1) electric flux are bound to lines on which $\tau^z \sigma^z =-1$, then confining the U(1) electric flux also eliminates two of the four possible edge types in the $\mathbb{Z}_2 \times \mathbb{Z}_2$ Walker-Wang model, leaving the $\mathbb{Z}_2^f$ fermionic Toric code, as desired.

\section{Statistics} \label{StatsApp}

In this Appendix we will derive the various claims made about the statistics of our point and line-like excitations made at various points in the main text.  

\subsection{Statistics in the bulk}
We begin with statistics in the bulk, and explicitly drop all boundary terms in our action (we return to these in Appendix \ref{SurfaceStatsApp}).  It is most convenient to separate the gauge field into static classical background fields, which we will call $A_\mu^{(0)}$, and the usual fluctuating parts.  We will assume that the static background fields $E_0$ and $B_0$ satisfy the classical equations of motion, such that the effective gauge field action in the bulk is
\ba
\mac{L} &=&  \frac{1}{4 g^2} \left( ( F_{\mu \nu}^{(0)})^2+ F_{\mu \nu} ^2 \right )  + \frac{2  \pi^2}{2 g^2} s_{\mu \nu} ^2 +i  \frac{\theta}{8}  \epsilon^{\mu \nu \rho \lambda} s_{\mu \nu} s_{\rho \lambda} \n
&&+\left( F_{\mu \nu } + F_{\mu \nu}^{(0)}  \right )  \frac{2  \pi}{2 g^2} s^{\mu \nu} -   i \left( A_\mu  + A_\mu^{(0)} \right) J_\mu  
\ea
As usual, 
\be
J_\mu = n_\mu + \frac{ \theta}{2 \pi} m_\mu
\ee 
is the total electric current, and we have integrated the $\theta$ term by parts.  

We can now integrate out the fluctuating gauge field $A$, to obtain:
\be \label{LineStats}
\mac{L} =  \frac{1}{4 g^2}  ( F_{\mu \nu}^{(0)})^2  + F_{\mu \nu }^{(0)}  \frac{2  \pi}{ g^2} s^{\mu \nu} -   i  A_\mu^{(0)} J_\mu  + \mac{L}_{\text{Coulomb}}
\ee
where
\begin{widetext}
\ba \label{Lcoul}
 \mac{L}_{\text{Coulomb}} & =& \frac{g^2}{2} J_\mu(r) G_{\mu \nu}(r-r')  J_\nu(r')  + \frac{ 2 \pi^2}{g^2} \partial^\mu s_{\alpha \mu} (r)  G^{\alpha \beta}(r-r')  \partial^\nu s_{\beta \nu}(r')  + \frac{  \pi^2}{ g^2} s_{\mu \nu}(r)  ^2  \n
 && +  2 \pi i \partial_\mu s_{\alpha \mu} (r) G^{\alpha \beta}(r-r')  J_\beta (r')  +
 i \theta \epsilon_{\beta \nu \rho \lambda } \left( \partial_\mu s_{\alpha \mu}(r)  G^{\alpha \beta}(r-r')  \partial_\nu s_{\rho \lambda} (r') + \frac{1}{4}  s_{\beta \nu}(r) s_{\rho \lambda}(r)\right) \n
 \ea
 \end{widetext}
 (Here we will work in imaginary time, where the statistical terms appear with explicit factors of $i$ in the action).  
There are two types of statistical interactions we will be interested in: those between point particles and the classical background gauge fields, and the statistical interactions between the various point particles, which are contained in $ \mac{L}_{\text{Coulomb}} $.  

\subsubsection{Statistics between point particles and static gauge fluxes}

The Berry phase between a charged particle and an external magnetic field can be read off from Eq. (\ref{LineStats}) in the usual way: we take
\be
J_\mu({\bf r}, t) = (\delta( {\bf r} -  {\bf r}_t ) ,  \delta( {\bf r} -{\bf r}_t) \sin t, - \delta( {\bf r} - {\bf r}_t)\cos t, 0 )
\ee  
where ${\bf r}_t =  (\cos t, \sin t, 0 )$, and $0 < t \leq 2 \pi$.  Taking $A_0 =0$, the net phase accumulated as the charge encircles this current loop is:
\be
\int_0^{2 \pi} dt 
\left( \bf{ A}^{(0)} ( {\bf r}_t , t)  \times r_t  \right) \cdot \hat{z}  
\ee
This is the usual loop integral $\oint {\bf A} \cdot {\bf dl}$, which simply gives the enclosed magnetic flux.  The important point to note is that this Berry phase depends on the total charge, and thus is $\theta$ dependent.  

Since the magnetic charge is unaffected by $\theta$, the Berry phase between a magnetic charge and an external electric field is necessarily $\theta$ independent, as can be shown directly from  Eq. (\ref{LineStats}). 

Note, however, that since for general $\theta$ the monopole carries both electric and magnetic charge, the Berry phase of a monopole around the flux tube created by another monopole is still $\theta$-independent. Consider a tube of electric flux $\frac{ \theta }{2 \pi}$ and magnetic flux $2 \pi$.  A monopole encircling this flux tube will acquire a Berry phase of $ \theta $ due to its electric charge ($q_E = \frac{ \theta }{2 \pi})$ encircling the magnetic flux, and an additional Berry phase of $- \theta $ due to its magnetic charge encircling the electric flux. Thus the statistics that we have found between point particles and static flux lines are compatible with the observation that statistics between point particles are $\theta$ independent. 

\subsubsection{Statistics between point particles}

The possible bulk statistical terms arising in $\mac{L}_{\text{Coulomb}}$ have been discussed, for example, in Ref. \cite{Goldhaber89}.  The terms in the first line of Eq. \ref{Lcoul} are (in the uncondensed phase) the usual Coulomb repulsion (attraction) between pairs of like (unlike) electric or magnetic charges.  The terms in the second line represent the possible statistical interactions.  

If we take $G_{\mu \nu} (q) = g_{\mu \nu} \frac{1}{q^2}$, which is the usual photon propagator in Feynman gauge, one can show\cite{Goldhaber89} that the two $\theta$-dependent terms on the second line cancel.  This leaves only the first term, which is a $\theta$ independent Berry phase interaction between charges and monopoles.  
%
 (Though these statistical calculations are carried out using the gauge field propagator appropriate to the Coulomb phase, their results remain valid for those charges (whether electric or magnetic) that cannot be screened by the condensate).\cite{Propitius95} 
For our purposes, this simply means\cite{Cardy82a} that a condensate of $(n, m)$ excitations will confine all point particles $(n', m')$ for which $n m' \neq n' m$.

\subsection{Statistics at the surface} \label{SurfaceStatsApp}

We now turn to the somewhat more delicate question of statistics at the surface.  This is discussed in detail in Ref. \onlinecite{Metlitski13} for the case $\theta = 2 \pi$ and in the Coulomb phase; here we will give the obvious generalization for arbitrary $\theta$.  Again, we will assume that the statistical interactions between charges that cannot be screened are identical in the Coulomb and confining phases.  

In Eq. \ref{LineStats}, we have dropped the following surface terms:
\ba \label{StatEq1}
\mac{L}_{\text{surf}} &=& \frac{\theta}{8 \pi^2}  \int _{\partial M}d^3 r_{\parallel} \epsilon^{\mu \nu \rho} \left \{  A_{\mu}^{(0)} \left(  \partial_\nu A^{(0)}_\rho - 2 \pi s_{\nu \rho}^{(0)} \right) \right . \n
&& \left. + A_{\mu} \left(  \partial_\nu A_\rho - 2 \pi s_{\nu \rho} \right)  \right \}
\ea
where $\mu \nu \rho$ are directions ``in'' the boundary, which we will for convenience take to be $(x,y,t)$.   

Metlitski, Kane and Fisher\cite{Metlitski13} give a detailed account of how to integrate out the fluctuating gauge fields to derive the contribution of the surface term to the point particles' mutual statistics, in the case that the Dirac strings of monopoles do not cross the surface (i.e. $s_{\mu \nu}=0$ in Eq. \ref{StatEq1}).   To do so, it is convenient to specify the type of boundary in question; we will consider the boundary between a region with $\theta = 2 \pi k$ and one with $\theta =0$.  Following Metlitski et al.'s approach, for two sources very close (relative to their separation) to the boundary, one obtains the total statistical interaction:
\be
\hat{\Theta}_{ij}  \epsilon_{\nu \lambda \sigma } \int d^4 r d^4 r'
 \epsilon_{\mu \nu \rho }\partial^{\mu } J^{(i)}_\rho (r)   \partial^\lambda  G_3( r_\parallel - r'_\parallel) \epsilon_{ \alpha \sigma \beta } \partial^\alpha J^{(j)}_\beta(r') 
\ee
where all indices are in the space-time directions of the boundary, and $r_{\parallel}$ indicates the separation along the direction of the boundary.  Here $G_3$ obeys
\be
\partial^2 G_3(r-r') = \delta(r-r')
\ee
The total statistical matrix, including both bulk and surface contributions, is: 
\be \label{ThetaEq}
\hat{\Theta} = \frac{\pi}{1 + k^2 \alpha^2}  \begin{pmatrix} -k  \alpha^2 & 1/2 & -1/2 \\
1/2 & k/4 & -k/4 \\
-1/2 &- k/4 & k/4  + k( 1 + k^2 \alpha^2) \\
\end{pmatrix}
\ee
Following Ref. \cite{Metlitski13}, we have expressed this matrix in the basis 
\be
\int_{- \infty}^0 d x^3 ( J(x^\mu), m_+(x^\mu), m^{(0)}_- (x^\mu) )
\ee
where $m_+$ is a monopole just above the surface, in the region where $\theta =0$, $m^{(0)}_-$ is a {\it neutral} monopole in the region just below the surface, where $\theta = 2 \pi k$, and $J$ is the total charge current (including both the boson current $n^\mu$ and the induced charge current of the monopole).  The term $k \pi$ in $\hat{\Theta}_{3 3}$ stems from the fact that if $k$ is odd, this neutral monopole is a fermion.   In the basis used in Eq. (\ref{ThetaEq}), the bare monopole current in the region with non-trivial $\theta$ is represented by the vector $(k, 0,1)$.  The objects $(0,1,0)$ (a monopole outside the bulk) and $(k,0,1)$ (a monopole inside the bulk, with its induced charge) have the same statistics for any $k$ and $g$.

In the obliquely confined phase, at long wavelengths the coupling will flow to $g^2 \rightarrow \infty$.   In this case, we obtain
\be
\hat{\Theta} =  \begin{pmatrix} -\frac{\pi}{k}  & 0 & 0 \\
0 & 0 &0 \\
0 &0 &   \pi k  \\
\end{pmatrix}\punc{.}
\ee
Hence in the 
confined phase, the charges have a statistical interaction proportional to $-\pi/k$, and a  charge near the surface (whether bare or induced) has no $\pi$ Berry phase with a monopole coming from above or below the surface.  The monopoles in the region $\theta = 2 \pi k$ have statistics of $k \pi$, indicating that they are fermionic of $k$ is odd and bosonic otherwise.

One might worry that allowing monopole world-lines to cross the surface will alter the statistics in such a way that the pure charge $(1,0,0)$ becomes confined if Dirac strings crossing the surface proliferate.  Physically, this could only arise if the actual Dirac string (rather than the monopole itself, whose statistics we have explicitly calculated here) became physically observable when it crossed the surface.  Such a change is expected at fractional $k$, where we expect the effective radius of compactification of our U(1) gauge field to change at the surface.  For integer $k$, however, this is not expected and one would not anticipate that including Dirac strings that cross the surface would alter our evaluation of the statistics.

\subsection{A more heuristic picture of the surface} \label{heuristic surface}

\begin{figure}[ht]
\includegraphics[width=.9\linewidth]{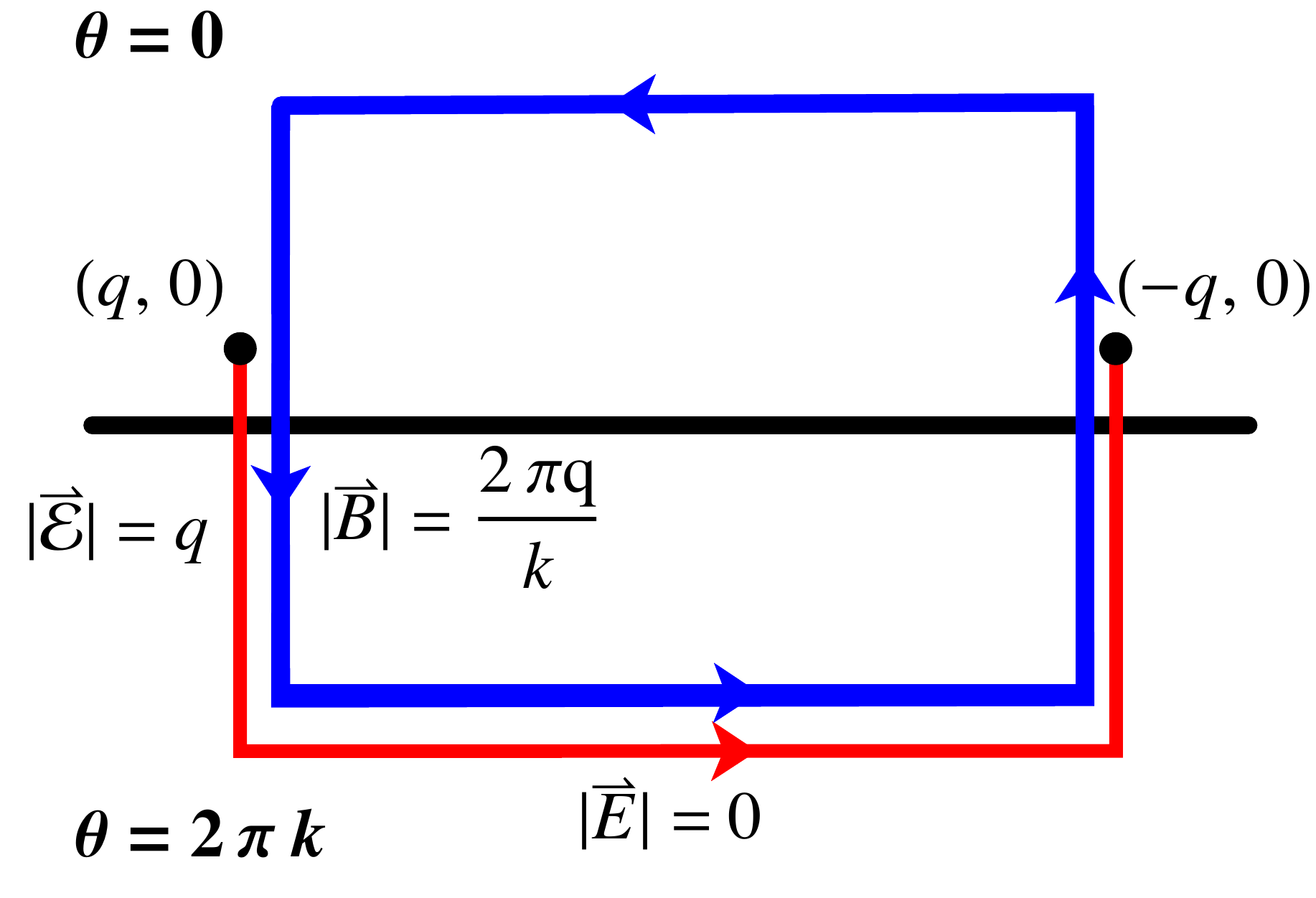}\\
\begin{centering}
\caption{(Color online): This figure shows the interface (black) between the $\theta=2\pi k$ dyon condensed phase, and the $\theta=0$ confined phase. The black dots are charge $q=1,\ldots,p-1$ test charges near the interface. Electric flux $E_{i}=\mathcal{E}_{i}-\frac{\theta}{4\pi^2} B_{i}$ is linearly confined in both phases. As a result, the $q$ charges are dressed with a $2\pi q/k$ magnetic flux tube (blue). The red lines denote $\mathcal{E}_{i}$ flux, defined below \eqnref{eq:Eflux}.}
\label{fig:surfacedeconf} 
\par\end{centering}
\end{figure}

One of the defining features of the Walker-Wang models that we consider (with the exception of the fermionic toric code) is that they have excitations with anyonic statistics pinned to their surfaces. How do these anyonic excitations arise in the gauge theory? The correspondence that we have established between axion electrodynamics and Walker-Wang Hamiltonians suggests that the `smooth' Walker-Wang boundary\cite{vonkeyserlingk13a} is equivalent to the interface between the dyon condensed phases at $\theta = 2 \pi k, \theta =0$. We can support this claim by examining the effective Lagrangian \eqnref{eq:bFbbJ} at large $g^{2}$ 

\begin{displaymath}
   \mathcal{L}= \left\{
     \begin{array}{ll}
        -\frac{i\pi}{k}b\wedge b- ib \wedge F 		&: \theta =  2\pi k \neq 0 \\
       +\frac{g^{2}}{2}b_{\mu \nu}^2- ib \wedge F  &: \theta =0 
     \end{array}
   \right.
\end{displaymath} 
where $b\in\{0,1,\ldots, p-1\}$ and $F\in \frac{2\pi}{p} \{0,1,\ldots,p-1\}$. The $bF+bb$ field theory arising for $\theta =  2\pi k \neq 0$ we showed (rigorously in some cases) is the same as a U$(1)_k$ Walker-Wang model. These models have ground states involving a superposition of loops of all lengths, encoded by $b$. The second case $\theta =0$ represents a trivial phase with no (or short) loops (i.e., $b=0$). Thus it is plausible that a boundary between the $\theta=2\pi k$ and $\theta=0$ large $g^{2}$ phases is akin to a boundary between a Walker-Wang model and a trivial (no loop) phase. Such a boundary condition is reminiscent of the `smooth' boundary in Ref.~\onlinecite{vonkeyserlingk13a}, for which the Walker-Wang models under consideration admit deconfined surface anyons. 

Here we argue heuristically for the existence of these deconfined anyons starting from the continuum action \eqnref{eq:effaction}. This supplements the more concrete field theoretic companion calculation in the previous subsection. To understand the deconfinement in the continuum setting, it is useful to consider what happens when we bring an integer charge $(q,0)$ with $q=1,2,\dots,p-1$  near the surface, just outside of the $\theta=2\pi k$ region (\figref{fig:surfacedeconf}). Such an object will source a field line $\mathcal{E}=q$. Outside of the sample, this corresponds to an electric $E=q$ field line which is linearly confined. Inside of the sample, this corresponds to an electric field value of $E=q-\frac{k}{2\pi} B$. Thus, within the sample, the field line need not be confined provided there is a nearby parallel magnetic field line $B= 2\pi q/k$. We do not require that $k$ divides $q$, so generically this magnetic field line cannot terminate on a monopole. It will form a closed loop shown in blue in \figref{fig:surfacedeconf}. A loop of purely magnetic flux and $\mathcal{E}=0$ is confined in the $\theta=2\pi k$ bulk (because $E=\mathcal{E}-\frac{k}{2\pi} B$), but not within the $\theta=0$ bulk. Thus, a pair of $(q,0)$ charges are not confined near the surface because the $\mathcal{E}$ field line (red) connecting $(q,0)$ charges can be screened by a magnetic flux loop as shown in \figref{fig:surfacedeconf}. 

The statistics of point excitations near the surface follow readily from this flux attachment picture. Surface $(q,0)$ particles are associated with closed $B= 2\pi q/k$ flux tubes, part of which lie in the $\theta=2\pi k$ region as shown in \figref{fig:surfacedeconf}. The exchange of any two $(q,0)$ excitations is thus associated with the intersection of two $\frac{2\pi q}{k}$ flux tubes in the  $\theta=2\pi k$ region. However such an intersection of flux tubes is associated with a phase of $ \pi q^2 /k$, by virtue of the $\frac{i k}{4\pi}F\wedge F$ term. However, there is also a phase of $-2\pi q^2/k$ due to the Berry phases of the $q$ charges with the $2\pi q/k$ magnetic flux tubes. Thus, the $(q,0)$ particles behave like anyons with a self-statistic of $e^{-i\pi q^2/k}$, in agreement with the result in the previous subsection.

\end{document}